\documentclass[12pt]{article}
\usepackage{amsmath,amssymb,amscd}
\usepackage{array}
\def\thickone{{\rm 1\mskip-4.5mu l}}
\usepackage[colorlinks=true,linkcolor=blue,urlcolor=blue]{hyperref}
\usepackage[T1]{fontenc}
\def\tempcite#1{\hbox{[\quad]}}
\newcommand{\ps}{(\sigma)}
\newcommand{\eps}{\epsilon}
\newcommand{\cjs}{c_j\ps}
\newcommand{\hcjs}{\hat{c}_j\ps}
\newcommand{\hdz}{\hat{\delta}_{0j}\ps^{(26-d)}}
\newcommand{\djs}{D_j\ps}
\newcommand{\uts}{U(1)^{26}}
\newcommand{\hj}{\mskip4mu\hat{\mskip-3mu\text{\it\j}}\mskip2mu}
\newcommand{\hJ}{\hat{J}}
\newcommand{\bhj}{\mskip4mu\bar{\hat{\mskip-3mu\text{\it\j}}}\mskip2mu}
\newcommand{\fjs}{f_j\ps}
\newcommand{\fls}{f_{\ell}\ps}
\newcommand{\FLs}{F_L\ps}
\newcommand{\op}{\phantom{x}\!\!\!\!^{'}}
\newcommand{\hatjovfs}{\mathchoice
{\tfrac{\hj }{f_j\ps}}
{\frac{\hj }{f_j\ps}}
{\frac{\hj }{f_j\ps}}
{\tfrac{\hj }{f_j\ps}}}
\newcommand{\hatlovfs}{\mathchoice
{\tfrac{\hat{\ell}}{f_\ell\ps}}
{\frac{\hat{\ell}}{f_\ell\ps}}
{\frac{\hat{\ell}}{f_\ell\ps}}
{\frac{\hat{\ell}}{f_\ell\ps}}}
\newcommand{\nrovrhos}{\mathchoice
{\tfrac{n(r)}{\rho\ps}}
{\frac{n(r)}{\rho\ps}}
{\frac{n(r)}{\rho\ps}}
{\frac{n(r)}{\rho\ps}}}
\newcommand{\nsovrhos}{\mathchoice
{\tfrac{n(s)}{\rho\ps}}
{\frac{n(s)}{\rho\ps}}
{\frac{n(s)}{\rho\ps}}
{\frac{n(s)}{\rho\ps}}}
\newcommand{\hatlovfjs}{\mathchoice
{\tfrac{\hat{\ell}}{f_j\ps}}
{\frac{\hat{\ell}}{f_j\ps}}
{\frac{\hat{\ell}}{f_j\ps}}
{\frac{\hat{\ell}}{f_j\ps}}}
\newcommand{\hell}{\hat{\ell}}
\newcommand{\hatLovfs}{\mathchoice
{\tfrac{\hat{L}}{F_L\ps}}
{\frac{\hat{L}}{F_L\ps}}
{\frac{\hat{L}}{F_L\ps}}
{\frac{\hat{L}}{F_L\ps}}}
\newcommand{\hatJovfs}{\mathchoice
{\tfrac{\hat{J}}{F_J\ps}}
{\frac{\hat{J}}{F_J\ps}}
{\frac{\hat{J}}{F_J\ps}}
{\frac{\hat{J}}{F_J\ps}}}

\newcommand{\jeaj}{J_{\eps a j}}
\newcommand{\jllj}{J_{\hat{L} L j}}
\newcommand{\jlll}{J_{\hat{L} L \ell}}
\newcommand{\jjjj}{J_{\hat{J} J j}}
\newcommand{\kd}[2]{\delta_{#1,#2}}
\newcommand{\ket}[1]{\mathop{{|}\mskip1mu #1\mskip1mu\rangle}\nolimits}
\newcommand{\kets}[2]{\mathop{{|}\mskip1mu #1\mskip1mu\rangle}\nolimits\mskip-4mu_{#2}}
\newcommand{\Z}{\mathbb{Z}}

\newcommand{\nosub}[2]{\:{:}\mskip1mu #1{:}_{#2}}
\newcommand{\lb}[1]{\label{#1}}
\def\cramped#1{\radical0 {\kern-\nulldelimiterspace#1}}
\newcommand{\Hp}{H({\rm perm})_{26-d}^{'}}

\def\mfrac#1#2{{\textstyle{\textstyle\medmuskip1mu\hbox{$#1$}\lower2.5pt\hbox{}\over
      \textstyle\medmuskip1mu\hbox{$\cramped{#2}$}\vbox to9pt{}}}}
\newcommand{\sS}[2]{_{#1}^{\smash{\phantom{#1}}#2}}
\let\tsty\textstyle

\def\babs#1{\bigl| #1\bigr|}

\def\bnorm#1{\bigl\| \,#1\,\bigr\|}
\def\BIgl({\mathopen{\hbox{\smaller${\biggl(}$}}}
\def\BIgr){\mathclose{\hbox{\smaller${\biggr)}$}}}
\def\bil({\mathopen{\raise.25pt\hbox{\smaller${\bigl(}$}}}
\def\bir){\mathclose{\raise.25pt\hbox{\smaller${\bigr)}$}}}

\numberwithin{equation}{section}

\makeatletter

\def\env@tcases{%
  \let\@ifnextchar\new@ifnextchar
  \left\lbrace
  \def\arraystretch{1.0}%
  \array{@{}l@{\quad}l@{}}%
  \noalign{\vskip-4pt}
}
\makeatother

\makeatletter

\newcommand{\Rmnum}[1]{\expandafter\@slowromancap\romannumeral #1@}
\makeatother

\title{The Orbifold-String Theories of Permutation-Type:\\ 
III. Lorentzian and Euclidean Space-Times\\
 in a Large Example}
\author{M.~B. Halpern\thanks{halpern@physics.berkeley.edu}\\
Department of Physics\\
University of California, 
Berkeley, California 94720, USA}
\date{}

\begin{document}

\maketitle

\begin{abstract}
\noindent To illustrate the general results of the previous paper, we 
discuss here a large concrete example of the orbifold-string theories of permutation-type.
For each of the many subexamples, we focus on evaluation of the 
\emph{target space-time dimension} $\hat{D}_j(\sigma)$,  
the \emph{target space-time signature} and the \emph{target space-time symmetry} 
of each cycle $j$ in each twisted sector $\sigma$. We find in particular a gratifying 
\emph{space-time symmetry enhancement} which naturally matches the space-time 
symmetry of each cycle to its space-time dimension. Although the orbifolds of $\Z_2$-permutation-type are 
naturally Lorentzian, we find that the target space-times associated to larger permutation
groups can be Lorentzian, Euclidean and even null 
($\hat{D}_j(\sigma)=0)$, with
varying space-time dimensions, signature and symmetry in a single orbifold.
\end{abstract}

\clearpage
\tableofcontents
\clearpage
\section{Introduction}\label{Section 1}
% MANU 1.1

In the present series of papers [1,2], we are studying the algebraic 
formulation of the general bosonic prototype
of the new orbifold-string theories of permutation-type, emphasizing the \emph{target space-times} of these theories.
The general prototypes include the string theories [3]
\label{1.1}
\begin{equation}
	\frac{{\rm U}(1)^{26 K}}{H_+},\quad \left[\frac{{\rm U}(1)^{26K}}{H_+}\right]_{{\rm open}} , 
	\thickspace H_+\subset H({\rm perm})_K \times H'_{26}
\end{equation}
thus extending our previous work [4-7] which was primarily at $K=2$.  
Here, the permutation group $H({\rm perm})_K$ permutes the $K$ copies of the critical closed string ${\rm U}(1)^{26}$, while $H_{26}'$
(the untwisted space-time automorphism group) is any automorphism group which acts uniformly on each copy.  
The closed-string (generalized permutation) orbifolds ${\rm U}(1)^{26 K}/ H_+$ involve left- and right-mover copies of the
algebras of the open-string counterparts $[{\rm U}(1)^{26 K}/ H_+]_{{\rm open}}$. 
  
Our description also includes
the orientation-orbifold string systems [3-6]:
\label{1.2}
\begin{equation}
	\frac{{\rm U}(1)^{26}}{H_-} = \frac{{\rm U}(1)_L^{26}\times{\rm U}(1)_R^{26}}{H_-}, 
	\quad H_-\subset \Z_2({\rm w.s.}) \times H'_{26}.
\end{equation}
In these cases $\Z_2({\rm w.s.})$ permutes the left-and right-movers of the 
critical closed string, and these string systems contain an equal number of twisted open- and closed-string sectors. The
open-string sectors of the orientation-orbifolds are included
in $[{\rm U}(1)^{52}/ H_+]_{{\rm open}}$, while the closed-string sectors 
form the space-time orbifold $U(1)^{26}/H'_{26}$.

We remind that these theories have many sectors $\{\sigma\}$ 
corresponding to the equivalence classes of the divisors $H_{\pm}$, so 
that each $\sigma$ corresponds to a choice of element in the permutation group $H({\rm 
perm})_{K}$ or $\Z_2({\rm w.s.})$ as well as the space-time automorphism group $H'_{26}$.

%MANU 1.1'

With the help of BRST quantization [4,1], the conformal cycle-dynamics of the general bosonic prototype
has been determined [2] up to the choice of the automorphism subgroup 
$H_{26}'$ and, so that the paper is essentially self-contained,  we begin here with
a brief review of this dynamics.
%these?

%MANU 1.1''

In particular, we know the extended physical-state conditions and orbifold Virasoro generators
for each cycle $j$ of each sector $\sigma$ of the open-string prototype in Eq.~(1.1)
%MANU 1.2
\begin{subequations}
\label{1.3}
	\begin{gather}
	\label{1.3a}
	\begin{gathered}[b]
	\bigl( \hat{L}_{\hj j} ((m+\hatjovfs) \geq 0) - \hat{a}_{\fjs} \kd{m+\hatjovfs}{\:0}\bigr) \kets{\chi\ps}{j} = 0 \\
	\end{gathered}\\
	\label{1.3b}
	\begin{gathered}[b] 
	\hat{L}_{\hj j} (m+\hatjovfs)=\tfrac{1}{2\fjs}\sum_{n(r)\mu\nu} \mathcal{G}^{n(r)\mu; -n(r),\, \nu}\ps
	\sum_{\hat{\ell}=0}^{\fjs-1}\sum_{p\in\Z} \times\\
	\hskip8em{} \times \negthinspace \nosub{\hat{J}_{n(r)\mu\hat{\ell} j}\bigl(p+\nrovrhos +\hatlovfjs\bigr)
	 \hat{J}_{-n(r),\nu,\hj-\hat{\ell},j}\bigl(m-p-\nrovrhos +\tfrac{\hj-\hat{\ell}}{\fjs}\bigr)}{M}\\
	 + \; \delta_{m+\hatjovfs,\, 0}\; \hat{\Delta}_{0j}\ps
	 \end{gathered}\\
	 \label{1.3c}
	 \begin{gathered}[b]
	 \hat{a}_{\fjs} = \tfrac{13 f^2_j\ps - 1}{12 \fjs}
	 \end{gathered}
	 \end{gather}
\end{subequations}
where $\hat{a}_{\fjs}$ in Eqs.~(1.3a,c) is the \emph{intercept} of cycle 
$j$ in sector $\sigma$. The associated mode algebras are as follows
%manu 1.3
\begin{subequations}
\label{1.4}
\begin{gather}
\label{1.4a}
\begin{gathered}[b]
\bigl[ \hat{L}_{\hj j}(m+\hatjovfs),\hat{L}_{\hat{\ell}\ell}(n+\hatlovfs)\bigr]\hfill\\
\qquad{}
= \delta_{j\ell }\Bigl\{
(m-n+\tfrac{\hj-\hat{\ell}}{\fjs}) \hat{L}_{\hj+\hat{\ell},j}(m+n+\tfrac{\hj+\hat{\ell}}{\fjs})\hfill\\
\hskip8em{}
+\tfrac{26\fjs}{12}(m+\hatjovfs)\bigl((m+\hatjovfs)^2 - 1\bigr)
\kd{m+n+\frac{\hj+\hat{\ell}}{\fjs}}{0}\Bigr\},
\end{gathered}
\\
\label{1.4b}
\begin{gathered}[b]
\bigl[
\hat{L}_{\hj j}(m+\hatjovfs), \hat{J}_{n(r)\mu \hat{\ell}\ell}(n+\nrovrhos +\hatlovfs)\bigr]\hfill\\
\hskip6em{}
= -\delta_{j\ell}(n+\nrovrhos  + \tfrac{\hat{\ell}}{\fls}) \hat{J}_{n(r)\mu,\, \hj+\hat{\ell},\, j}
(m+n+\nrovrhos +\tfrac{\hj+\hat{\ell}}{\fjs}),
\end{gathered}
\\
\label{1.4c}
\begin{gathered}[b]
\bigl[
\hat{J}_{n(r)\mu\hj j}(m+\nrovrhos + \hatjovfs), 
\hat{J}_{n(s)\nu\hat{\ell}\ell}(n+\tfrac{n(s)}{\rho\ps} + \hatlovfs)\bigr]\hfill\\
\hskip3em{}
=\delta_{j\ell }\fjs (m +\nrovrhos + \hatjovfs) \kd{n(r)+n(s)}{\,0\bmod\rho\ps}\hfill\\
\hskip13em{}
\times \kd{m+n+\frac{n(r)+n(s)}{\rho\ps}+\frac{\hj+\hat{\ell}}{\fjs}}{\,0}
\, \mathcal{G}_{n(r)\mu;-n(r),\, \nu}\ps,
\end{gathered}
\\
\label{1.4d}
\begin{gathered}[b]
\hat{L}_{\hj \pm\fjs,j}(m+\tfrac{\hj\pm\fjs}{\fjs}) = \hat{L}_{\hj j}(m\pm 1 + \hatjovfs),
\end{gathered}
\\
\label{1.4e}
\begin{split}
\hat{J}_{n(r)\pm \rho\ps,\mu\hj j}(m +\tfrac{n(r)\pm\rho\ps}{\rho\ps} + \hatjovfs) & =
\hat{J}_{n(r)\mu,\hj\pm \fjs,j}(m+\nrovrhos +\tfrac{\hj\pm\fjs}{\fjs})\\
& =
\hat{J}_{n(r)\mu\hj j}(m\pm 1 + \nrovrhos +\hatjovfs),
\end{split}
\\
\label{1.4f}
\bhj = 0,1,\dots, \fjs - 1, \quad
j = 0,1,\dots, N\ps - 1, \quad
{\textstyle\sum\limits_j} \fjs = K
\end{gather}
\end{subequations}\\
where Eq.~(1.4a) is a set of orbifold Virasoro algebras [8,16,1,2].

The results in Eqs.~(1.3) and (1.4) are the description of the theories at \emph{cycle central charge} $\hat{c}_j\ps=26\fjs$, where $\fjs$ is the length of cycle $j$ in
each element $\sigma$ of $H({\rm perm})_K$ or $\Z_2({\rm w.s.})$. The 
sector central charges $\hat{c}(\sigma)=26K$ of each string theory are obtained by summing over 
the cycles with the sum rule in Eq.~(1.4f). The reader is reminded that the 
(closed-string) generalized permutation orbifolds in Eq. (1.1) additionally require a corresponding set of right-mover operators.
We will return below to specify the normal-ordering of the orbifold 
Virasoro generators, the precise forms of the conformal 
weights $\{\hat{\Delta}_{0j}(\sigma)\}$ and in particular the quantities 
$\{n(r)\mu, \rho(\sigma),\mathcal{G}(\sigma)\}$ -- 
which together encode the specific element $\omega(\sigma)$ of the space-time 
symmetry group $H'_{26}$.

%manu 1.4
We also know a map [2] from the $\hat{c}_j\ps=26\fjs$ description of the extended physical-state condition
(1.3) of each cycle $j$ to an \emph{ equivalent,reduced} formulation of 
the physical states of cycle $j$ at \emph {reduced cycle central 
charge} $c_j\ps=26$.  The explicit form of the map between the operators of the two 
formulations is 
%manu1.4'
\begin{subequations}
\label{1.5}
\begin{gather}
\label{1.5a}
L_j(M_j) \equiv 
\fjs \hat{L}_{\hj j}(m+\hatjovfs) - \tfrac{13}{12}(\fjs^2 -1)\kd{m+\hatjovfs}{0}
\\
\label{1.5b}
J_{n(r)\mu j}(M_j + \fjs\tfrac{n(r)}{\rho\ps})\equiv
\hat{J}_{n(r)\mu \hj j}(m+\nrovrhos +\hatjovfs),
\\
\label{1.5c}
M_j \equiv \fjs m + \bhj, \quad \bhj = 0,1,\dots, \fjs - 1
\end{gather}
\end{subequations}
%manu1.5
%\newpage
\noindent where $M_j\in\Z$ covers the integers once 
%manu1.5'
\footnote{The existence of the map (1.5) and the resulting physical-state 
description (1.6a) of cycle $j$ at $c_j\ps=26$
has colorfully been termed "black magic" by our old friend J.-B. Zuber.  But in fact, the map is only
a straightforward generalization of the inverse of the so-called "orbifold induction procedure" [8], which was
a foundational step of the orbifold program [8-22].
Examples of the map were also given in Ref.~[5].}. Then the reduced (unhatted) operators provide the following 
equivalent, reduced description of cycle $j$ in sector $\sigma$:
%manu1.5''
%To illustrate the general results of the previous paper, we consider here
 %a large, explicit set of examples of the new orbifold-string theories of permutation type,
  %using both the description at central charge $\hat{c}_j\ps=26\fjs$ and the equivalent, reduced formulation
   %at $c_j\ps=26$, we focus in particular on the target \emph{space-time dimension} $\hat{D}_j \ps$ of cycle $j$ in
   % each twisted sector $\sigma$ of these theories, including the signature and \emph{space-time symmetry} of each cycle.
%where the physical states here are the same states, now re-expressed in 
%terms of the reduced currents, as those in    
\begin{subequations}\label{1.6}
\begin{gather}
\label{1.6a}
\bigl(L_j (M\geq 0) -\kd{M}{0}\bigr)
\kets{\chi\ps}{j} = 0\\[0.7em]
\label{1.6b}
\begin{gathered}
L_j(M) = \delta_{M,0}\,\hat{\delta}_{0j}\ps
+ \tfrac{1}{2}\sum_{n(r)\mu\nu} \mathcal{G}^{n(r)\mu;\,-n(r) \nu}\ps
\sum_{P\in\,\Z}\, \times\\
\hskip5em{}\times
\nosub{J_{n(r)\mu j}\bigl(P + \fjs\nrovrhos \bigr)\,J_{-n(r),\nu j}\bigl(M-P-\fjs\nrovrhos\bigr)}{M}.
\end{gathered}
\end{gather}
\end{subequations}
%Note  that the reduced physical-state conditions (1.6a) exhibit unit 
%intercept, as in ordinary untwisted string theory. We emphasize that the physical states 
%described by the reduced conditions are exactly the same physical states as those 
%described in Eq.~(1.3a), now re-expressed via the map in terms of the 
%reduced currents. The precise form of the conformal-weight shifts 
%$\{\hat{\delta}_{0j}\ps\}$ will also be reviewed below. 
The precise form of the conformal-weight shifts 
$\{\hat{\delta}_{0j}\ps\}$ in Eq.~(1.6b) will also be reviewed below. The mode algebras of the  reduced formulation are as follows
%manu1.6
\begin{subequations}\label{1.7}
\begin{gather}
\label{1.7a}
[L_j(M),L_\ell(N)] = \delta_{j\ell}
\{ (M-N)L_j(M+N) + \tfrac{26}{12} M(M^2 -1)\kd{M+N}{0}\},\\[0.7em]
\label{1.7b}
\begin{gathered}[b]
\hskip-16em[L_j(M), J_{n(r)\mu \ell}(N+ f_\ell\ps\nrovrhos) ]\\
\qquad= -\delta_{j\ell}(N+\fjs \nrovrhos)J_{n(r)\mu \ell}(M+N+\fjs\tfrac{n(r)}{\rho\ps}),
\end{gathered}\\[0.7em]
\label{1.7c}
\begin{gathered}[b]
[J_{n(r)\mu j}(M+ \fjs\nrovrhos), J_{n(s)\nu \ell}(N+ f_\ell\ps\nsovrhos)]\hfill\\
\hskip4em{}=
\delta_{j\ell}(M+\fjs \nrovrhos)\,\kd{n(r)+n(s)}{0\bmod\rho\ps}\,
\kd{M+N+\fjs\smash[t]{\tfrac{n(r)+n(s)}{\rho\ps}}}{\,0}\\
%{\scriptstyle \bullet} 
\times \;\mathcal{G}_{n(r)\mu;-n(r),\nu}\ps,
\end{gathered}
\\[0.7em]
\label{1.7d}
J_{n(r)\pm \rho\ps,\mu j}\bigl(M+\fjs\tfrac{n(r)\pm\rho\ps}{\rho\ps}\bigr)=
J_{n(r)\mu j}\bigl(M\pm\fjs+\fjs\tfrac{n(r)}{\rho\ps}\bigr).
\end{gather}
\end{subequations}
%manu1.7
We see in particular that the reduced Virasoro generators $L_j(M)$ satisfy the ordinary 
Virasoro algebras (1.7a) at reduced cycle central charges 
$c_{j}(\sigma)=26$, and we note as well that the reduced physical-state 
condition (1.6a) of each cycle 
exhibits unit intercept, as in ordinary untwisted string theory. It must 
be emphasized [5,2] that the physical states 
described by the reduced conditions are exactly the same physical states as those 
described in Eq.~(1.3a), now re-expressed via the map in terms of the 
reduced currents.
% The precise form of the conformal-weight shifts 
%$\{\hat{\delta}_{0j}\ps\}$ will also be reviewed below. 

In the reduced description, one has unwound the part of the twist associated to cycle $j$ of the relevant element of $H({\rm perm})_K$, 
leaving unconventionally-twisted currents with fractional modeing $(f_{j}(\sigma)n(r))/\rho(\sigma)$.  We know that the unreduced 
formulation is local [3], so one suspects 
that the equivalent, reduced formulation is generically non-local, a question to which we shall return in later papers.  An exception to this
intuition of course is the case $\fjs=1$ where, in agreement with the map, the reduced formulation is the same as the unreduced.

%manu1.8
We turn now to the precise forms [2] of various structures in the formulations above. In both the original and reduced
formulations, the orbifold Virasoro generators (1.3b) and the reduced 
Virasoro generators (1.6b) are defined as mode-normal-ordered forms:
\begin{equation}
\label{1.8}
\nosub{A(\xi)B(\eta)}{M}
\equiv
\theta(\xi\geqslant0)B(\eta)A(\xi)+\theta(\xi<0)A(\xi)B(\eta).
\end{equation}
The conformal weights $\hat{\Delta}_{0j}\ps$ and conformal weight-shifts $\hat{\delta}_{0j}\ps$ of cycle $j$ in sector $\sigma$ have
been computed as follows: 
\begin{subequations}\label{1.9}
\begin{gather}
\label{1.9a}
\hat{\Delta}_{0j} \ps = \tfrac{13}{12}\Bigl(\fjs - \mfrac{1}{\fjs}\Bigr) + \mfrac{1}{\fjs}\hat{\delta}_{0j}\ps,\\[0.7em]
\label{1.9b}
\hat{\delta}_{0j}\ps \equiv \tfrac{f_{j}(\sigma)}{2} \sum\limits_r \dim [\bar{n}(r)] \, \hat{A} [\tfrac{\bar{n}(r)}{\rho\ps}],\\
\label{1.9c}
\hat{A} [\tfrac{\bar{n}(r)}{\rho\ps}] \equiv
\tfrac{\fjs}{2} \sum\limits_{\hj = 0}^{\fjs - 1} 
 \bigl(\tfrac{\bar{n}(r)}{\rho\ps} - \tfrac{\hj}{\fjs}\bigr)
\bigl(\tfrac{\hj+1}{\fjs} - \tfrac{\bar{n}(r)}{\rho\ps}\bigr)\,
\theta\bigl(\hatjovfs \leqslant \tfrac{\bar{n}(r)}{\rho\ps} < \tfrac{\hj+1}{\fjs}\bigr).
%%bigger parentheses 
%%\Bigl(\tfrac{\bar{n}(r)}{\rho\ps} - \tfrac{\hj}{\fjs}\Bigr)
%%\Bigl(\tfrac{\hj+1}{\fjs} - \tfrac{\bar{n}(r)}{\rho\ps}\Bigr)\,
%%\theta\Bigl(\hatjovfs \leqslant \tfrac{\bar{n}(r)}{\rho\ps} < \tfrac{\hj+1}{\fjs}\Bigr).
\end{gather}
\end{subequations}
Other properties of the function $\hat{A} [\frac{\bar{n}(r)}{\rho\ps}]$ are given in Ref.~[2].

%manu1.9
Finally, the choice of element $\omega\ps \in H_{26}' $ is encoded in the new string theories by first solving
the eigenvalue problem [10,12] of $\omega\ps$
\begin{subequations}\label{1.10}
\begin{gather}
\label{1.10a}
\omega( \sigma) \sS{a}{b}\, U^\dagger\ps\sS{b}{n(r)\mu} = 
U^\dagger \ps\sS{a}{n(r)\mu} e^{-2\pi i \nrovrhos }\\[0.4em]
\label{1.10b}
\omega( \sigma) \sS{a}{c}\, \omega( \sigma) \sS{b}{d} \,G_{cd} = G_{ab}, \qquad 
G = \Bigl(\hskip-2pt \begin{array}{rc}-1 & 0 \\  0 & \thickone \end{array} \Bigr),\\[0.4em]
\label{1.10c}
a,b = 0,1,\dots,25, \qquad
\bar{n}(r)\in (0,1,\dots,\rho\ps - 1),\\[0.7em]
\label{1.10d}
{\tsty\sum\limits_\mu} = \dim [\bar{n} (r)], \quad
{\tsty\sum\limits_r} \dim [\bar{n} (r)] = 26
\end{gather}
\end{subequations}
where $G$ is the untwisted target-space metric of $\uts$ and $n(r),\mu$ 
are called respectively the spectral and degeneracy indices of the element.  The quantity 
$\rho\ps$ is the order of $\omega\ps$, and $\bar{n}(r)$ is the pullback to the fundamental region
of the spectral indices. This information appears in the modeing of both 
formulations, where we also see the twisted metric $\mathcal{G}_{\scriptscriptstyle{\bullet}}\ps$
%The explicit choice of $\omega\ps\in H_{26}'$ is then encoded in the unreduced
%and reduced formulations
%by the quantities $\{n(r)\mu, \rho(\sigma)\}$ as well as the \emph{twisted metric} $\mathcal{G}_{\scriptscriptstyle{\bullet}}\ps$
and its inverse $\mathcal{G}^{\scriptscriptstyle{\bullet}}\ps$:
\begin{subequations}
\label{1.11}
\begin{gather}
\label{1.11a}
\begin{split}
\mathcal{G}_{n(r)\mu ; \,n(s)\nu} \ps &= \chi_{n(r)\mu} \ps \, \chi_{n(s)\nu}\ps \,
U\ps\sS{n(r)\mu}{a} \,U\ps\sS{n(s)\nu}{b} \,G_{ab}\\
&= 
\kd{n(r) + n(s)}{0 \bmod \rho\ps}\,\mathcal{G}_{n(r)\mu ; -n(r),\nu}\ps,
\end{split}\\[0.7em]
\label{1.11b}
\begin{split}
\mathcal{G}^{n(r)\mu ; n(s)\nu} \ps &= 
\chi_{n(r)\mu}^{-1}\ps \, \chi_{n(s)\nu}^{-1}\ps \,G^{ab}\,U^\dagger\ps\sS{a}{n(r)\mu}\, U^\dagger\ps\sS{b}{n(s)\nu}\\
&=
\kd{n(r) + n(s)}{0 \bmod \rho\ps}\,\mathcal{G}^{n(r)\mu ; -n(r),\nu} \ps,
\end{split}
\\[0.7em]
%manu1.9'
\label{1.11c}
\sum_{n(t)\eta} \mathcal{G}^{n(r)\mu;\,n(t)\eta}\ps \,\mathcal{G}_{n(t)\eta ; \,n(s)\nu} \ps = \delta \sS{n(r)\mu}{n(s)\nu}.
\end{gather}
\end{subequations}
The quantities $\{\chi_{n(r)\mu}\ps\}$ are a set of essentially arbitrary normalizations.
Our first task in the body of this paper will be to choose a large, explicit set of automorphism groups $H_{26}'$, so
that the eigenvalue problems and hence the operator systems in either formulation can be fully evaluated.

%manu1.10
A central question in the physics of the new string theories is the 
\emph{target space-time} of each cycle $j$ in sector $\sigma$,
which we have discussed in general [2], without choosing any particular non-trivial $H_{26}'$.  In fact,
all salient features of the target space-times are \emph{invariant} under the choice of unreduced or reduced formulation.  
Thus we have seen that [2]
%manu1.11
\begin{subequations}\label{1.12}
\begin{gather}
\label{1.12a}
\hat{D}_j \ps =\dim \{ \hat{J}_j (0)_{\sigma} \}=D_{j}\ps= \dim\{{J}_j (0)_{\sigma}\},\\[0.4em]
\label{1.12b}
\hat{P}_j^2 \ps = P_j^2 \ps, \\[0.4em]
\label{1.12c}
\hat{P}_j^2 \ps_{(0)} = P_j^2 \ps_{(0)}= 2 (\hat{\delta}_{0j}\ps -1)\geqslant -2,\\[0.4em]
\label{1.12d}
\Delta (\hat{P}_j^2 \ps) = 2\fjs \babs{ m +\tfrac{\bar{n}(r)}{\rho(\sigma)} +\tfrac{\bhj }{f_j\ps}}\\[0.4em]
\label{1.12e}
\,\,\,=\Delta (P_j^2 \ps) = 2\,\babs{M +\fjs\tfrac{\bar{n}(r)}{\rho(\sigma)}},\\[0.4em]
\label{1.12f}
\hat{P}_j^2 \ps_{(0)}^{^{\scriptstyle \rm closed}} = P_j^2 \ps_{(0)}^{^{\scriptstyle \rm closed}}= 4 (\hat{\delta}_{0j}\ps -1)\geqslant -4
\end{gather}
\end{subequations}
%manu1.12
where $\{\hat{J}_j (0)_{\sigma}=\hat{J}_j (0)_{\sigma}\}$ are the zero modes and $\hat{D}_{j}(\sigma)=D_{j}(\sigma)$ is the \emph{target space-time dimension} of 
cycle $j$ in sector $\sigma$. The general forms of the momentum-squared 
operators (1.12b) are given
in Ref.~[2], and the quantities $P_j^2 \ps_{(0)}$ and $P_j^2 \ps^{\;\;{\rm closed}}_{(0)}$ are the ground-state mass-squareds
respectively of the open- and closed- string sectors. Finally, the level-spacings $\Delta (P_j^2 \ps)$ are the 
increments of $P_j^2 \ps$ associated
 to adding negative-moded currents to the states of cycle $j$.
  %manu1.5''
%To illustrate the general results above, we will consider here
%a large, explicit set of examples of the new orbifold-string theories of permutation type,
%using both the description at central charge $\hat{c}_j\ps=26\fjs$ and the equivalent, reduced formulation
%at $c_j\ps=26$. We focus in particular on the target \emph{space-time dimension} $\hat{D}_j \ps$ of cycle $j$ in
%each twisted sector $\sigma$ of these theories, including the signature and \emph{space-time symmetry} of each cycle.  
%In these examples we will find many simple orbifold-string theories where the space-times are entirely Lorentzian, but 
%higher sub-examples can be Lorentzian,
%Euclidean, and even $\textrm{null} (\hat{D}_j \ps=~\mskip-7mu0)$, with varying space-times even in a single orbifold.

To illustrate the general results above, we will study in this paper
a large, explicit set of examples of the new orbifold-string theories of permutation-type,
using both the description at cycle central charge $\hat{c}_j\ps=26\fjs$ and the equivalent, reduced formulation
at reduced cycle central charge $c_j\ps=26$. We will focus here on the computation of the target space-time 
dimension $\hat{D}_j \ps$ of each cycle $j$ in
every sector $\sigma$ of these theories (see Secs.~4 and 5), including the 
\emph{target space-time signature} and the \emph{target-space-time 
symmetry} of each cycle (see Secs.~7, 8 and 10).

In particular, we will use the simplicity of the equivalent, reduced 
formulation to see a gratifying \emph{target space-time symmetry 
enhancement} (again see Secs.~7,8 and 10) which naturally matches the full space-time symmetry of each 
cycle $j$ to the space-time dimensions of the cycle. We will find many simple orbifold-string theories
(for example the orbifolds of $\Z_2$-permutation-type) where the space-times are entirely Lorentzian, but 
higher sub-examples can be Lorentzian $(SO(\hat{D}_{j}(\sigma)-1,1))$,
Euclidean $(SO(\hat{D}_{j}(\sigma))$ and even $\textrm{null}\, (\hat{D}_j 
\ps=~\mskip-7mu0)$, with varying sector- and cycle-dependent space-times even in a single orbifold.

%manu2.1
\section{A Large Set of Examples}\label{Section 2}
%To illustrate the general results above, we will study in this paper
%a large, explicit set of examples of the new orbifold-string theories of permutation-type,
%using both the description at central charge $\hat{c}_j\ps=26\fjs$ and the equivalent, reduced formulation
%at $c_j\ps=26$. We focus in particular on the \emph{target space-time dimension} $\hat{D}_j \ps$ of cycle $j$ in
%each twisted sector $\sigma$ of these theories, including the 
%\emph{signature} and the residual \emph{space-time symmetry} of each cycle.  
%In these examples we will find many simple orbifold-string theories where the space-times are entirely Lorentzian, but 
%higher sub-examples can be Lorentzian,
%Euclidean, and even $\textrm{null}\, (\hat{D}_j \ps=~\mskip-7mu0)$, with varying space-times even in a single orbifold.

In the summary above, the subgroup $H_{26}'$ can be any automorphism group of the critical closed string $\uts$.  Here we begin
our discussion of a large set of examples with the following explicit choice of $H_{26}'$ :
\begin{subequations}
\label{2.1}
\begin{gather}
\label{2.1a}
H_{26}' \subset (\pm \thickone)_d \times \Hp\;,\\
\label{2.1b}
H_+\subset H({\rm perm})_K \times H_{26}' .
\end{gather}
\end{subequations}
The examples (2.1) then involve \emph{two} permutation groups, the basic permutation group $H({\rm perm})_K$ on $K$ copies
of $\uts$ and the spatial permutation group $H(\rm perm)_{26-d}'$ on $26-d$ spatial dimensions of $\uts$.
The orientation-orbifold string systems can be included with the same  $H'_{26}$ by the substitutions 
$H({\rm perm})_{2}\rightarrow \Z_{2}(\rm w.s.)$ and 
$H_{+}\rightarrow H_{-}$. Indeed,  examples of all the orbifold-string systems of 
$\Z_{2}$-permutation-type [5] have been studied with this choice of $H'_{26}$.
%These choices of $H'_{26}$ 
%therefore include all the subexamples of $\Z_{2}$-permutation-type studied in Ref.~[5]. 
%with $H(\rm perm)_2=\Z_2$.

%manu2.1'
More precisely, we will study all the sectors $\{\sigma\}$ associated to the elements of $H_{26}'$ 
\begin{subequations}
\label{2.2}
\begin{gather}\label{2.2a}
\omega\ps=
\begin{cases}
(\omega)_d=(\pm\thickone)_d \quad {\rm on} \quad a=0,1,\dots,d-1\,,\\
(\omega)_{26-d}\ps\in H({\rm perm})^{'}_{26-d} \quad  {\rm on} \quad a=d,\dots,25\,,
\end{cases}\\[0.7em]
\label{2.2b}
1\leqslant d \leqslant 26
\end{gather}
\end{subequations}
at each fixed j-cycle length $f_{j}(\sigma)$.

In these examples, the parameter $d$ partitions the 26 dimensions of each 
critical closed string $U(1)^{26}$ into a set of $d$ and $(26-d)$ dimensions.
Referring to the eigenvalue problem (1.10), we see that the action on the set of $d$ dimensions is very simple:
\begin{subequations}\label{2.3}
\begin{gather}
\begin{align}
(\omega)_d &= (\thickone)_d,\;\rho=1,\;\bar{n}=0,\\
(\omega)_d &= (-\thickone)_d, \;\rho = 2,\;\bar{n}=1,
\end{align}\\[0.7em]
\frac{\bar{n}}{\rho\ps}=\frac{\epsilon}{2}\,, \quad \epsilon=
\begin{cases}
0 \quad {\rm for} (\omega)_d = (\thickone)_d\,,\\
1 \quad {\rm for} (\omega)_d = (-\thickone)_d\,,
\end{cases}\\ \label{}
\mathcal{G}_{\scriptscriptstyle{\bullet}}\ps=G_{ab}^{(d)}, \quad 
\mathcal{G}^{\scriptscriptstyle{\bullet}}\ps=G_{(d)}^{ab},\quad a,b,\dots,d-1.
\end{gather}
\end{subequations}
Here we have chosen trivial eigenmatrices and $\mu=a$, as well as trivial 
normalizations $\{\chi\}$. The quantity
\begin{equation}\label{2.4}
G^{(d)}=\Bigl(\hskip-2pt\begin{array}{rc}-1 & 0 \\ 0 & \thickone \end{array}\Bigr)_d
\end{equation}
is the restriction of the 26-dimensional metric $G$ to the first $d$ dimensions.

%manu2.1''
The action of $H_{26}'$ on the remaining set of $(26-d)$ spatial 
dimensions is any element $(\omega)_{26-d}\ps$ of any permutation group on $(26-d)$ elements. 
The results for general permutation groups [14,16,3-5] are well known in the orbifold 
program and we find in these cases:
\begin{subequations}\label{2.5}
\begin{gather}
\label{a}
\frac{n(r)}{\rho\ps}=\frac{\hat{J}}{F_{J}\ps}\\[0.6em]
\label{2.5b}
\mathcal{G}_{\scriptscriptstyle{\bullet}}\ps = \mathcal{G}_{\hat{J}J;\,\hat{L}L}\ps=\delta_{JL}\,F_{J}\ps \,\delta_{\hat{J}+\hat{L},\,0\bmod F_J\ps}\\[0.7em]
\label{2.5c}
\mathcal{G}^{\scriptscriptstyle{\bullet}}\ps = \mathcal{G}^{\hat{J}J;\,\hat{L}L}\ps=\delta^{JL}\,\tfrac{1}{F_{J}\ps} \,\delta_{\hat{J}+\hat{L},\,0\bmod F_J\ps}\\[0.7em]
\label{2.5d}
\bar{\hat{J}}=0,1,\dots,F_J\ps-1\,,\;\, J=0,1,\dots,N\ps'-1\,,\;\, \sum_{J}F_J\ps=26-d\,.
\end{gather}
\end{subequations}
%manu2.2
Here $F_J\ps$ is the length of cycle $J$ in the element of $H({\rm perm})_{26-d}^{'}$, and
the reader may profit by comparing the ranges (2.5d) for the indices 
associated to the elements of $H({\rm perm})_{26-d}^{'}$ with
the ranges (1.4f) for the indices associated with the elements of the 
basic permutation group $H({\rm perm})_K$.  
The notation for our large example here is very similar to that  chosen in Ref.~[5], except that we
are using capital letters $\{\hat{J},J\}$ here for the elements of $H({\rm perm})_{26-d}^{'}$, in order to distinguish it from the
small letters $\{\hat{j},j\}$ we are now using to describe the elements of $H({\rm perm})_K$.

%manu2.3
Let us note with Eq.~(2.2b) that we cannot choose $d=0$ in the partition -- because $H({\rm perm})_{26}'$ is
not an automorphism group of the untwisted Lorentzian critical closed string $\uts$.
%manu2.4
Using in particular the twisted metrics (2.3d) and (2.5b,c) , we may then write down the explicit form
in our large example of the general results summarized in Sec.\ref{Section 1}.

%manu2.5
We begin with the orbifold Virasoro generators at $\hat{c}_j\ps=26\fjs$, 
which now take the following explicit form:
%manu2.5'
\begin{subequations}
\label{2.6}
\begin{gather}
\label{2.6a}
\begin{gathered}[b] 
\hat{L}_{\hj j} (m+\hatjovfs)=\, \delta_{m+\hatjovfs,\, 0}\; \hat{\Delta}_{0j}\ps\hfill\\ 
\hskip1em{}+\tfrac{1}{2\fjs} \,G_{(d)}^{ab}\, \negthinspace 
\nosub{\hat{J}_{\epsilon a \hat{\ell} j}(p+\tfrac{\epsilon}{2} +\hatlovfjs) 
\hat{J}_{-\epsilon,\,b,\,\hj-\hat{\ell},\,j}(m-p-\tfrac{\epsilon}{2} +\tfrac{\hj-\hat{\ell}}{\fjs})}{M}\\
\hskip-10em{}+\tfrac{1}{2\fjs}\sum_{L}\tfrac{1}{F_L\ps} \sum_{\hat{\ell}=0}^{\fjs-1}\sum_{\hat{L}=0}^{F_L\ps-1} \times\\
\hskip6em{} \times \negthinspace \nosub{\hat{J}_{\hat{L} L \,\hat{\ell} j}(p+\tfrac{\hat{L}}{F_L\ps} +\hatlovfjs) 
\hat{J}_{-\hat{L},\,L,\,\hj-\hat{\ell},\,j}(m-p-\tfrac{\hat{L}}{F_L\ps} +\tfrac{\hj-\hat{\ell}}{\fjs})}{M},
\end{gathered}\\[0.7em]
\label{2.6b}
\epsilon=
\begin{cases}
1 \quad {\rm for} (\omega)_d = (\thickone)_d\,,\\
0 \quad {\rm for} (\omega)_d = (-\thickone)_d\,.
\end{cases}
\end{gather}
\end{subequations}
Note that the orbifold Virasoro generators are additive with respect to the contributions
of $(\omega)_d=\pm(\thickone)_d$ and $\omega_{26-d}(\sigma) \in H({\rm 
perm})_{26-d}^{'}$. In what follows, we shall refer to these 
contributions respectively as those of type (d) and of type (26-d). 

Similarly, we find
the following explicit forms of the conformal weights and conformal-weight shifts:
%manu2.6
\begin{subequations}
\label{2.7}
\begin{gather}
\label{2.7a}
	\hat{\Delta}_{0j}\ps=\tfrac{13}{12}(\fjs-\tfrac{1}{\fjs})+\tfrac{1}{f_{j}(\sigma)}\hat{\delta}_{0j}\ps,\\[0.7em]
\label{2.7b}
	\hat{\delta}_{0j}\ps=\hat{\delta}_{0j}^{\,(d)}\ps+\hat{\delta}_{0j}^{\,(26-d)}\ps,\\[0.7em]
\label{2.7c}
	\hat{\delta}_{0j}^{\,(d)}\ps
	=\tfrac{d}{4}\sum_{\hj=0}^{\fjs-1}\bigl( \tfrac{\fjs\epsilon}{2}-\hj\bigr)\bigl(\hj+1-\tfrac{\fjs\epsilon}{2}\bigr)\,
	\theta\bigl((\tfrac{\fjs\epsilon}{2}-1)<\hj\leqslant\tfrac{\fjs\epsilon}{2}\bigr)\,\,\geqslant 0,\\[0.7em]
\label{2.7d}
\begin{gathered}[b]
	\hat{\delta}_{0j}^{\,(26-d)}\ps=\tfrac{1}{4}\sum_L\sum_{\hat{L}=0}^{F_L\ps-1}\sum_{\hj=0}^{\fjs-1}
	\bigl(\tfrac{\fjs\hat{L}}{F_L\ps}-\hj\bigr)\bigl(\hj+1-\tfrac{\fjs\hat{L}}{F_L\ps}\bigr) \times\\
	\hskip3em \times \; \theta\bigl((\tfrac{\fjs\hat{L}}{F_L\ps}-1)<\hj\leqslant\tfrac{\fjs\hat{L}}{F_L\ps}\bigr)\,\,\geqslant 0.
\end{gathered}
\end{gather}
\end{subequations}
Here the conformal-weight shifts are given as the sum of the 
contributions of type (d) and type (26-d), and we have used slightly rearranged arguments of the Heaviside functions $\theta$. 

%manu2.7
The corresponding twisted mode algebras include Eq.~(1.4a) for the orbifold Virasoro generators and the following algebras
involving the two types of twisted currents
%manu2.8
\begin{subequations}
\label{2.8}
\begin{gather}
\label{2.8a}
\begin{gathered}[b]
\hskip-4em{}\bigl[\hat{L}_{\hj j}(m+\hatjovfs), 
\hat{J}_{\eps a \hat{\ell}\ell} (n+\tfrac{\eps}{2} +\hatlovfs)\bigr]\hfill\\
\hskip3em{} = -\delta_{j\ell}(n+\tfrac{\eps}{2} + \hatlovfjs) \,
\hat{J}_{\eps a,\,\hj+\hat{\ell},\,j}(m+n+\tfrac{\eps}{2} +\tfrac{\hj+\hat{\ell}}{\fjs}),
\end{gathered}\\[0.7em]
\label{2.8b}
\begin{gathered}[b]
\bigl[\hat{J}_{\eps a \hj j}(m+\tfrac{\eps}{2} + \hatjovfs), 
\hat{J}_{\eps' b \hat{\ell} \ell }(n+\tfrac{\eps'}{2} +\hatlovfs)\bigr]\hfill\\
\hskip3em{}
=\delta_{j\ell }\fjs (m +\tfrac{\eps}{2} + \hatjovfs)\,G_{ab}^{(d)} \,\kd{\eps+\eps'}{\,0\bmod 2 \phantom{\tfrac{1}{2}}} \!\!\kd{m+n+\frac{\eps+\eps'}{2}+\frac{\hj+\hat{\ell}}{\fjs}}{\,0}\;\,,
\end{gathered}\\[0.7em]
\label{2.8c}
a,b=0,1,\dots,d-1,\\[0.7em]
\label{2.8d}
\begin{gathered}[b]
\hskip-2em{}\bigl[\hat{L}_{\hj j}(m+\hatjovfs), 
\hat{J}_{\hat{J}J\hat{\ell}\ell}(n+\tfrac{\hat{J}}{F_J\ps} +\hatlovfs)\bigr]\hfill\\
\hskip3em{} = -\delta_{j\ell}(n+\tfrac{\hat{J}}{F_J\ps} + \hatlovfjs) \,
\hat{J}_{\hat{J}J,\,\hj+\hat{\ell},\,j}(m+n+\tfrac{\hat{J}}{F_J\ps} +\tfrac{\hj+\hat{\ell}}{\fjs}),
\end{gathered}\\[0.7em]
\label{2.8e}
\begin{gathered}[b]
\hskip2em{}\bigl[\hat{J}_{\hat{J}J \hj j}(m+\tfrac{\hat{J}}{F_J\ps}  + \hatjovfs), 
\hat{J}_{\hat{L}L\hat{\ell}\ell}(n+\tfrac{\hat{L}}{F_L\ps} +\hatlovfs)\bigr]\hfill\\                                                                                                                                                                                                                                                                 
\hskip3em{}=\delta_{j\ell }\delta_{JL}\fjs F_J\ps(m +\tfrac{\hat{J}}{F_J\ps}  + \hatjovfs)\,
\kd{\hat{J}+\hat{L}}{\,0\bmod F_J\ps \phantom{\tfrac{1}{2}}}\!\! \kd{m+n+\frac{\hat{J}+\hat{L}}{F_J\ps}+\frac{\hj+\hat{\ell}}{\fjs}}{\,0}\;.
\end{gathered}
\end{gather}
\end{subequations}
%manu2.9
To these algebras we may append the periodicity relations (1.4d) of the 
orbifold Virasoro generators and the two types of currents:
\begin{subequations}
\label{2.9}
\begin{gather}
\label{2.9a}
\hat{J}_{\eps a,\hat{\ell}\pm f_\ell\ps,\ell}\bigl(m+\tfrac{\eps}{2}+\tfrac{\hat{\ell}\pm f_\ell\ps}{f_\ell\ps}\bigr)=
\hat{J}_{\eps a \hat{\ell}\ell}\bigl(m\pm 1+\tfrac{\eps}{2}+\tfrac{\hat{\ell}}{f_\ell\ps}\bigr),\\[0.7em]
\label{2.9b}
\hat{J}_{-\eps, a \hj j}\bigl(m-\tfrac{\eps}{2}+\tfrac{\hj}{\fjs}\bigr)=
\hat{J}_{\eps a \hj j}\bigl((m-\eps)+\tfrac{\eps}{2}+\tfrac{\hj}{\fjs}\bigr),\\[0.7em]
\label{2.9c}
\begin{gathered}[b]
\hat{J}_{\hat{L} L,\hat{\ell}\pm f_\ell\ps,\ell}\bigl(m+\tfrac{\hat{L}}{F_L\ps}+\tfrac{\hat{\ell}\pm f_\ell\ps}{f_\ell\ps}\bigr)=\hat{J}_{\hat{L}\pm F_L\ps,L \hat{\ell}\ell}\bigl(m+\tfrac{\hat{L}\pm F_L\ps}{F_L\ps}+\tfrac{\hat{\ell}}{f_\ell\ps}\bigr)\\
=\hat{J}_{\hat{L} L \hat{\ell} \ell}\bigl(m\pm1+\tfrac{\hat{L}}{F_L\ps}+\tfrac{\hat{\ell}}{f_\ell\ps}\bigr).
\end{gathered}
\end{gather}
\end{subequations}
Finally we add the \emph{adjoint operations} of the large example
\begin{subequations}
\label{2.10}
\begin{gather}
\label{2.10a}
\hat{L}_{\hj j}(m+\hatjovfs)^\dagger=\hat{L}_{-\hj, j}(-m-\hatjovfs),\\[0.7em]
\label{2.10b}
\hat{J}_{\eps a \hat{\ell} \ell}\bigl(m+\tfrac{\eps}{2}+\hatlovfs \bigr)^\dagger=\hat{J}_{-\eps, a, -\hat{\ell}, \ell}\bigl(m-\tfrac{\eps}{2}-\hatlovfs \bigr),\\[0.7em]
\label{2.10c}
\hat{J}_{\hat{L}L\hat{\ell}\ell}\bigl(m+\hatLovfs+\hatlovfs\bigr)^\dagger=\hat{J}_{-\hat{L},L,-\hell,\ell}\bigl(-m-\hatLovfs-\hatlovfs\bigr)
\end{gather}
\end{subequations}
which follow easily from the general orbifold adjoint operation of Ref.~[12].  Special cases of these adjoint operations
were given for the "pure" permutation orbifolds (trivial $H'_{26}$) in Refs.~[16,2].
%manu2.10

With the adjoint operations and the current-current commutators in Eqs.~(2.8b,e), we can compute the norm of any basis state
in cycle $j$ of sector $\sigma$, for example:
\begin{subequations}
\label{2.11}
\begin{gather}
\label{2.11a}
\begin{gathered}[b]
\hat{J}_{\eps a \hj j}\bigl((m+\tfrac{\eps}{2}+\tfrac{\hj}{\fjs})>0\bigr) \kets{0,\,\hat{J}_j(0)}{\sigma} \hfill\\
\hskip4em{}=\hat{J}_{\hat{J}J\hj j}\bigl((m+\tfrac{\hat{J}}{F_J\ps}+\tfrac{\hj}{\fjs})>0\bigr)\kets{0,\,\hat{J}_j(0)}{\sigma} = 0,
\end{gathered}\\[0.7em]
\label{2.11b}
\begin{gathered}[b]
\bnorm{\hat{J}_{\eps a \hell \ell}\bigl((m+\tfrac{\eps}{2}+\hatlovfs)<0\bigr) \kets{0,\,\hat{J}_j(0)}{\sigma}}^2\hfill\\
\hskip7em{}=G_{aa}^{(d)}\fls\bigl(m+\tfrac{\eps}{2}+\hatlovfs\bigr)\,\bnorm{\kets{0,\,\hat{J}_j(0)}{\sigma}}^2,
\end{gathered}\\[0.7em]
\label{2.11c}
\begin{gathered}[b]
\bnorm{\hat{J}_{\hat{L} L \hell \ell}\bigl((m+\hatLovfs+\hatlovfs)<0\bigr) \kets{0,\,\hat{J}_j(0)}{\sigma}}^2\hfill\\
\hskip7em{}=\fls\FLs\babs{m+\hatLovfs+\hatlovfs}\,\,\bnorm{\kets{0,\,\hat{J}_j(0)}{\sigma}}^2.
\end{gathered}
\end{gather}
\end{subequations}
Here $\kets{0,\,\hat{J}_j(0)}{\sigma}$ is the momentum-boosted 
twist-field state [2] of cycle $j$ in sector $\sigma$,
which is annihilated by all positively-moded currents.  The relation of 
this state to the physical ground-state of the cycle is discussed in 
Secs.~7-9.  Similarly, it is straightforward to compute the norms 
of any number of negatively-moded currents on this state and we conclude that the \emph{only} basis states with negative norm
are associated, as in ordinary string theory, with an odd number of time-like currents (i.e. $a=0$ because $G_{00}^{(d)}=-1$).
In particular, no negative norms are associated with the twisted currents $\{\hat{J}_{\hat{J}J\hj j}\}$ of the second type.

%manu2.11
To conclude the formulation of the large example at $\hcjs=26\fjs$, we remind the reader of 
the ground-state momentum-squared in Eq.~(1.12), and give the explicit 
form of the increments (level-spacing) of the momentum-squared
 %of cycle $j$
%in sector $\sigma$
\begin{subequations}
\label{2.12}
\begin{gather}
\label{2.12a}
\bigr(\hat{L}_{0j}(0)-\hat{a}_{\fjs}\bigl)\kets{\chi\ps}{j}=0,\\[0.7em]
\label{2.12b}
\hat{L}_{0j}(0)=\tfrac{1}{2\fjs}\bigl(-\hat{P}^2_j\ps+\hat{R}_{j}\ps\bigr)+\hat\Delta_{0j}\ps,\\[0.7em]
\label{2.12c}
\Delta(\hat{P}^2_j\ps)=\Delta(\hat{R}_{j}(\sigma))=
\begin{cases}
2\fjs \babs{m+\tfrac{\eps}{2}+\hatjovfs} \quad {\rm for} \quad \hat{J}\bigl((m+\tfrac{\eps}{2}+\hatjovfs)<0\bigr),\\[0.5em]
2\fjs \babs{m+\hatLovfs+\hatjovfs} \quad {\rm for} \quad \hat{J}\bigl((m+\hatLovfs+\hatjovfs)<0\bigr)
\end{cases}
\end{gather}
\end{subequations}
which are obtained on addition of a negatively-moded current of either type to a 
physical state. We shall return later to the fact
that all the increments are strictly positive. The explicit form of the 
generalized number operator 
$\hat{R}_{j}(\sigma)=\hat{R}_{j}(\sigma)^\dagger\geq 0$ is easily 
obtained from Eqs.~(2.6a),(2.10) and (2.12b).

%manu2.12
Next, we give the explicit form of the large example in the equivalent \emph{reduced} 
formulation at reduced cycle central charge $\cjs=26$.  
We begin with the explicit form of the reduced Virasoro generators
\begin{multline}\label{2.13}
\hskip4em{}L_j(M)=\, \delta_{M,\,0}\; \hat{\delta}_{0j}\ps\\[0.7em]
+\tfrac{1}{2} \,G_{(d)}^{ab} \sum_{P\in\Z} \negthinspace 
\nosub{J_{\eps a j}\bigl(P+\tfrac{\fjs\epsilon}{2}\bigr) 
\hat{J}_{-\eps,\,bj}\bigl(M-P-\tfrac{\fjs\epsilon}{2}\bigr)}{M}\\[0.5em]
+\tfrac{1}{2}\sum_{L}\tfrac{1}{F_L\ps} \sum_{\hat{L}=0}^{\FLs-1}\sum_{P\in\Z}
\nosub{J_{\hat{L}Lj}\bigl(P+\tfrac{\fjs\hat{L}}{\FLs}\bigr) 
\hat{J}_{-\hat{L},Lj}\bigl(M-P-\tfrac{\fjs\hat{L}}{\FLs}\bigr)}{M}
\end{multline}
where the conformal weight-shifts $\{\hat{\delta}_{0j}\ps\}$ of the large example are given in Eqs.~(2.7b-d).  The reduced mode algebras consist
of the ordinary Virasoro algebras (1.7a) at $\cjs=26$ and the algebras 
involving the reduced current modes of both types:
%manu2.13
\begin{subequations}\label{2.14}
\begin{gather}
\label{2.14a}
\hskip-4em{}\bigl[L_j(M), J_{\eps a \ell}(N+\tfrac{\fls\eps}{2})\bigr]
= -\delta_{j\ell}\bigl(N+\tfrac{\fls\eps}{2}\bigr) \,
\jeaj \bigl(M+N+\tfrac{\fjs\eps}{2}\bigr),\\[0.7em]
\label{2.14b}
\begin{gathered}[b]
\hskip-4.8em{}\bigl[J_{\eps aj}(M+\tfrac{\fjs\eps}{2}), 
J_{\eps' b \ell}(N+\tfrac{\fls\eps}{2})\bigr]\hfill\\[0.7em]
=\delta_{j\ell }\,G_{ab}^{(d)} \bigl(M+\tfrac{\fjs\eps}{2}\bigr)\,\kd{\eps+\eps'}{\,0\bmod 2 \phantom{\tfrac{1}{2}}} 
\!\!\kd{M+N+\fjs\tfrac{\eps+\eps'}{2}}{\,0}\;\,,
\end{gathered}\\[0.7em]
\label{2.14c}
\begin{gathered}[b]
\hskip-9em{}\bigl[L_j(M), \jlll(N+\tfrac{\fls\hat{L}}{\FLs})\bigr]\hfill\\[0.7em]
\hskip-4em{}= -\delta_{j\ell}\bigl(N+\tfrac{\fls\hat{L}}{\FLs}\bigr) \, \jllj \bigl(M+N+\tfrac{\fjs\hat{L}}{\FLs}\bigr),
\end{gathered}\\[0.7em]
\label{2.14d}
\begin{gathered}[b]
\hskip-3em{}\bigl[\jjjj(M+\tfrac{\fjs \hat{J}}{F_J\ps}), 
\jlll(N+\tfrac{\fls\hat{L}}{\FLs})\bigr]\hfill\\                                                                                                                                                                                                                                                                 
\hskip0em{}=\delta_{j\ell }\,\delta_{JL}\, F_J\ps\bigl(M +\tfrac{\fjs \hat{J}}{F_J\ps}\bigr)\,
\kd{\hat{J}+\hat{L}}{\,0\bmod F_J\ps \phantom{\tfrac{1}{2}}}\!\!\kd{M+N+\fjs\frac{\hat{J}+\hat{L}}{F_J\ps}}{\,0}\;.
\end{gathered}
\end{gather}
\end{subequations}
Similarly, the reduced periodicities and adjoint operations are obtained 
as follows
\begin{subequations}
\label{2.15}
\begin{gather}
\label{2.15a}
J_{-\eps,aj}(M-\tfrac{\fjs\eps}{2})=\jeaj((M-\fjs\eps)+\tfrac{\fjs\eps}{2}),\\
\label{2.15b}
J_{\hat{L}\pm\FLs,Lj}(M+\fjs\tfrac{\hat{L}\pm\FLs}{\FLs})=\jllj(M\pm\fjs+\fjs\hatLovfs),\\
\label{2.15c}
L_j(M)^\dagger=L_j(-M),\\
\label{2.15d}
\jeaj(M+\tfrac{\fjs\eps}{2})^\dagger=J_{-\eps,aj}(-M-\tfrac{\fjs\eps}{2}),\\
\label{2.15e}
\jllj(M+\tfrac{\fjs\hat{L}}{\FLs})^\dagger=J_{-\hat{L},Lj}(-M-\tfrac{\fjs\hat{L}}{\FLs}).
\end{gather}
\end{subequations}
%manu2.15
Then the reduced current-current commutators and adjoint operations directly
give the following reduced analogues of the norm computations: 
\begin{subequations}
\label{2.16}
\begin{gather}
\label{2.16a}
 \kets{0,\,J_j(0)}{\sigma} = \kets{0,\,\hat{J}_j(0)}{\sigma} \\[0.7em]
\label{2.16b}
\jeaj\bigl((M_j+\tfrac{\fjs \eps}{2})>0\bigr) \kets{0,\,J_j(0)}{\sigma} =\jjjj \bigl((M_j+\tfrac{\fjs \hat{L}}{\FLs})>0\bigr)\kets{0,\,J_j(0)}{\sigma} = 0,\\[0.7em]
\label{2.16c}
\begin{gathered}[b]
\bnorm{\jeaj\bigl((M_j+\tfrac{\fjs \eps}{2})<0\bigr) \kets{0,\,J_j(0)}{\sigma}}^2\hfill\\
\hskip7em{}=G_{aa}^{(d)}\,\babs{M_j+\tfrac{\fjs \eps}{2}}\;\bnorm{\kets{0,\,J_j(0)}{\sigma}}^2,
\end{gathered}\\[0.7em]
\label{2.16d}
\begin{gathered}[b]
\bnorm{\jllj\bigl((M_j+\tfrac{\fjs \hat{L}}{\FLs})<0\bigr) \kets{0,\,J_j(0)}{\sigma}}^2\hfill\\
\hskip7em{}=\FLs\, \babs{M_j+\tfrac{\fjs \hat{L}}{\FLs}}\;\bnorm{\kets{0,\,J_j(0)}{\sigma}}^2.
\end{gathered}
\end{gather}
\end{subequations}
With $M_j=\fjs m+\bhj$, the reduced norms are seen to be the same as the 
unreduced norms in Eq.~(2.11). Indeed, since the map (1.5) is only an operator-relabeling,
 all inner products  are the same in the two formulations.
In particular we see again with Eqs.~(2.14b) and (2.15d) that the only negative-norm basis states are those with an odd number of negatively-moded time-like currents
 $\{J_{\eps 0 j}\}$, in parallel to ordinary string theory.

%manu2.16
Finally, the increments of the reduced momentum-squared $P_j^2 \ps$ are obtained
\begin{subequations}
\label{2.17}
\begin{gather}
\label{2.17a}
(L_j(0)-1)\kets{\chi\ps}{j}=0,\quad \forall j,\sigma\\[0.7em]
\label{2.17b}
L_j(0)=\tfrac{1}{2}\bigl(-P^2_j\ps+R_{j}\ps\bigr)+\hat\delta_{0j}\ps\\[0.7em]
\label{2.17c}
\Delta(P^2_j\ps)=\Delta(R_{j}(\sigma))=
\begin{cases}
2 \babs{M_j+\tfrac{\fjs \eps}{2}} \quad {\rm for} \quad J\bigl((M_j+\tfrac{\fjs \eps}{2})<0\bigr)\\[0.5em]
2 \babs{M_j+\tfrac{\fjs \hat{L}}{\FLs}} \quad {\rm for} \quad J\bigl((M_j+\tfrac{\fjs \hat{L}}{\FLs})<0\bigr)
\end{cases}
\end{gather}
\end{subequations}
from the reduced physical-state conditions (1.6a) when a negatively-moded reduced current of either type is added to the state.
Using $M_j=\fjs m +\bhj$, it is easily seen that these increments are the 
same as the unreduced increments in Eq.~(2.12c). The explicit form of the 
reduced number operator 
$R_{j}(\sigma)=R_{j}(\sigma)^\dagger=\hat{R}_{j}(\sigma)\geq 0$ can be 
obtained from Eqs.~(2.13),(2.15) and (2.17b).

%manu3.1
\section{Target Space-Time Dimensions}\label{Section 3}
In this and the following two sections, we will focus on locating and counting the \emph{zero modes} of the currents, 
which define the \emph{momenta} of cycle $j$ in sector $\sigma$ and hence
the \emph{dimensions of the target space-time} in each cycle $j$.
These issues were discussed for the general case in Ref.~[2], emphasizing 
that the target space-time structure is invariant under the reduction, and we 
concentrate here on detailed analysis of the large example in the reduced formulation. 
%Preliminary discussion of these issues for the general orbifold-string 
%system of permutation-type was given in Ref.~[2], and we will
%concentrate here on detailed analysis of the large example, using  
%issues have been discussed for the general orbifold-string system in 
%Ref.~2, and we will For 
%this purpose, we will primarily employ the reduced
%formulation at $\cjs=26$, recalling that the zero modes are invariant under the reduction. 

Thus we know that [5,2] 
\begin{subequations}\label{3.1}
\begin{gather}
\label{3.1a}
\{\hJ_j(0)_\sigma\}=\{J_j(0)_\sigma\}, \quad \; \forall j,\sigma\\[0.5em]
\label{3.1b}
\bigl[ \{\hJ_j(0)_\sigma \}, \{\hJ_\ell(0)_\sigma \} \bigr]= 
\bigl[ \{J_j(0)_\sigma \}, \{J_\ell(0)_\sigma \} \bigr]=0, \quad \; \forall j,\ell,\sigma
\end{gather}
\end{subequations}
and indeed it is easily checked from the current-current commutators 
(2.8b.e) or (2.14b,d) that the zero-modes commute with the non-zero modes in each formulation. The
momentum-squared operators $\hat{P}^2_j\ps=P^2_j\ps$ (see Ref.~[2]) are
related to the orbifold Virasoro generators in Eqs.~(2.12b) and 
Eq.~(2.17b), and the explicit forms of these operators in the large example will be given below.

%manu3.2
It will be convenient in the analysis to consider the zero-modes of type $\!(d)$ and the zero modes of
type $\!(26-d)$ separately.  In particular, we shall obtain more explicit 
expressions for the \emph{target space-time dimension} of cycle $j$ in sector $\sigma$
\begin{subequations}\label{3.2}
\begin{gather}
\label{3.2a}
\hat{D}_j\ps = \dim \{\hJ_j(0)_\sigma \}=D_j\ps=\dim \{J_j(0)_\sigma \}\\[0.5em]
\label{3.2b}
= D_j\ps^{(d)}+D_j\ps^{(26-d)}.
%%\hskip-3.7em{}= D_j\ps^{(d)}+D_j\ps^{(26-d)}.
\end{gather}
\end{subequations}
Similarly, we will find the momentum-squared operators in the form
%manu3.3
\begin{equation}\label{3.3}
\hat{P}^2_j\ps=P^2_j\ps=P^2_j\ps^{(d)}+P^2_j\ps^{(26-d)}.
\end{equation}
Since they arise from the $(26-d)$ space-like dimensions of the ordinary closed string $\uts$, the
 momenta in $P^2_j\ps^{(26-d)}$ are expected to be spacelike ($P^2_j\ps^{(26-d)}\leqslant0$ in our metric), a fact that will be verified in Sec. 5.

%manu4.1
\section{The Zero Modes of Type \!($\mathbf{d}$)}\label{Section 4}

The zero modes of type $\!(d)$ are easily located in the reduced currents of type $\!(d)$
\begin{equation*}
\bigl\lbrace J_{\epsilon aj}(M_j+\tfrac{\fjs\eps}{2}), \quad a=0,1,\dots,d-1\bigr\rbrace
\end{equation*}
where we remind that the parameter $\eps=0,1$ corresponds respectively to 
the d-dimensional automorphisms $(\omega)_d=(\thickone)_d$ and $(-\thickone)_d$. 
The results are
\begin{subequations}\label{4.1}
\begin{gather}
\label{4.1a}
\bigl\lbrace J_{0aj}(0), \;\; a=0,1,\dots,d-1\bigr\rbrace \quad \textrm{for} \quad (\omega)_d=(\thickone)_d,\\[0.5em]
\label{4.1b}
\bigl\lbrace J_{1aj}(0), \;\; a=0,1,\dots,d-1\bigr\rbrace \quad 
\textrm{for} \quad (\omega)_d=(-\thickone)_d \;\, \textrm{and} \; \fjs=\textrm{even},\\[0.5em]
\label{4.1c}
\textrm{no zero modes of type (d) when}  \;(\omega)_d=(-\thickone)_d \;\, \textrm{and} \; \fjs=\textrm{odd}.
\end{gather}
\end{subequations}
The total number of zero modes of type $\!(d)$ in cycle $j$ of sector 
$\sigma$ is then summarized as follows: 
\begin{subequations}\lb{4.2}
\begin{gather}
\lb{4.2a}
\hskip-7em{}D_j\ps^{(d)}=\tfrac{d}{2}\bigl(1+(-1)^{\eps\fjs}\bigr)\\[0.5em]
\hskip5em{}=
\begin{cases}
d \quad {\rm for} \quad \eps=0\quad or\,\,\, \eps=1 \,\,\textrm{and} \;\fjs \;\textrm{even}\\[0.5em]
0 \quad {\rm for} \quad \eps=1 \;\, \textrm{and} \;\fjs \;\textrm{odd}.
\end{cases}
\end{gather}
\end{subequations}
The corresponding contribution of the zero modes of type $\!(d)$ to the momentum-squared operator of cycle $j$ in sector $\sigma$
%manu4.2
\begin{subequations}\lb{4.3}
\begin{gather}
\lb{4.3a}
\hat{P}^2_j\ps^{(d)}=P^2_j\ps^{(d)}=
\tfrac{1}{2}\bigl(1+(-1)^{\eps\fjs}\bigr)\,\eta_{(d)}^{ab}\jeaj(0)\,J_{-\eps,bj}(0)\,,\\[0.5em]
\lb{4.3b}
\eta_{(d)}\equiv -G_{(d)}=
\Bigl(\hskip-2pt \begin{array}{rc}1 & 0 \\  0 & -\thickone \end{array} \Bigr)_{(d)}
\end{gather}
\end{subequations}
also follows from the general results of Ref.~[2]. The currents with 
$(-\epsilon)$ are defined by the periodicity relation (2.9b).

%manu4.2'
Including then the adjoint operation in Eq.~(2.15d), we may further verify that
\begin{subequations}\lb{4.4}
\begin{gather}
\lb{4.4a}
P^2_j\ps^{(d)\dagger}=P^2_j\ps^{(d)},\\[0.5em]
\lb{4.4b}
P^2_j\ps^{(d)}=\tfrac{1}{2}\bigl(1+(-1)^{\eps\fjs}\bigr)
\sum_{a=0}^{d-1} \eta_{(d)}^{aa}\,\babs{\jeaj(0)}^2
\end{gather}
\end{subequations}
where the form  (4.4b) of the momentum-squared holds on any state with diagonalized momenta.

%manu4.3
To these observations, we add a closely-related result.  It is not difficult
to do the $\hj$ sum in Eq.~(2.7c), obtaining the following explicit form 
for the contribution of the currents of type $\!(d)$ to the conformal-weight shifts:
\begin{subequations}\lb{4.5}
\begin{gather}
\lb{4.5a}
\hskip-7em{}\hat{\delta}_{0j}\ps^{(d)}=\tfrac{d}{32}\bigl(1-(-1)^{\eps\fjs}\bigr)\\[0.5em]
\lb{4.5b}
\hskip5em{}=\begin{cases}
0 \quad {\rm for} \quad \eps=0\quad or\,\,\eps =1 \,\,\textrm{and} \;\fjs \;\textrm{even}\\[0.5em]
\tfrac{d}{16} \quad {\rm for} \quad \eps=1 \;\, \textrm{and} \;\fjs \;\textrm{odd}.
\end{cases}
\end{gather}
\end{subequations}
%In fact, for the first half of Eq.~(4.5b), no terms at all are allowed by 
%the Heaviside functions of (2.17c), while for the second half of (4.5b) only
In completing these sums, one notices for the first two cases of 
Eq.~(4.5b) that no terms at all are allowed by the Heaviside functions of 
Eq.~(2.7c). Similarly for the last case of (4.5b), only
the term with $\hj=(\fjs -1)/2$ contributes.
%$\hj=\tfrac{\fjs-1}{2}$ contributes.

It is instructive then to consider the results (4.5) together with the previous 
result (4.2) for the number of space-time dimensions of type $\!(d)$.  In particular consider the cases
$\eps=0$ or $\eps=1$ and $\fjs$ even, for which there are $d$ zero modes and no conformal-weight shifts of type $\!(d)$.  We can see
 both of these results explicitly
%manu4.4
in the contributions of type $\!(d)$ to the reduced Virasoro generators (2.13), which can be put in the form
\begin{equation*}
\kd{M}{0}\,(\hat{\delta}_{0j}\ps^{(d)}=0)-\tfrac{1}{2} \eta_{(d)}^{ab}\sum_{P\in\Z}
\nosub{\jeaj(P) \,J_{-\eps,bj}(M-P)}{M}
\end{equation*}
by shifting the integer $P$ in the sum.  In these cases, we see only integer-moded sequences which (as in the ordinary untwisted string)
can never [2] produce conformal-weight shifts.  In the other case ($\eps=1$ and $\fjs$ odd), we see only fractional-moded sequences
(no zero modes) and non-zero conformal-weight shifts.

%manu5.1
\section{The Zero Modes of Type \!($\mathbf{26-d}$)}\label{Section 5}
We consider next the zero modes of the currents of type (26-d), 
associated to each element $\omega\ps_{26-d}\in \Hp$:
\begin{subequations}\lb{5.1}
\begin{gather}
\lb{5.1a}
\jllj(M+\fjs\hatLovfs)\\[0.5em]
\lb{5.1b}
\hat{L}=0,1,\dots,\FLs-1\,,\quad L=0,1,\dots,N\ps{'}-1\,,\\[0.5em]
\lb{5.1c}
\FLs\geqslant 1\,,\quad\sum_L \FLs =26-d\,,\quad1\leqslant d\leqslant 26.
\end{gather}
\end{subequations}
We remind that $f_{j}(\sigma)$ is the length of cycle $j$ in the chosen 
element of $\rm {H(perm)_{K}}$, while $F_{L}(\sigma)$ and $N\ps^{'}$  are respectively the length of cycle $L$ 
and the number of $L$-cycles in the chosen element of $\Hp$.
%where $N\ps^{'}$ is the number of $L$-cycles in $\omega\ps_{26-d}$.  
This part of the problem is
trivial ($\hat{D}_j\ps^{(0)}=0$) for $d=26$ because the group $H({\rm perm})^{'}_0$ is trivial, with no room
for any $L$-cycles in $\sum_L \FLs=0$.  For the rest of this section we therefore restrict the discussion to the
non-trivial range $1\leqslant d\leqslant 25$.  (We shall return to include
the $d=26$ theories $H_+\subset H({\rm perm})_K \times (\pm \thickone)_{26}$ in the later discussion, reminding
the reader here only that the "pure" permutation orbifolds with $H_+\subset H({\rm perm})_K \times (\thickone)_{26}$ have already
been discussed in Ref.~[2]).

%manu5.2
The problem of finding the zero modes of the currents in Eq.~(5.1) is equivalent to finding all solutions of the conditions:
\begin{equation}\lb{5.2}
\hat{L}=\tfrac{\FLs\hj'}{\fjs}\,,\quad \hj'\in\{0,1,\dots,\fjs-1\}\,,\quad \hat{L}\in\{0,1,\dots,\FLs-1\}.
\end{equation}
These conditions are a special case of the general description of the zero-mode problems for all $H^{'}_{26}$ in Ref.~[2].

On examination of these conditions, we find that the solutions of the 
zero-mode problem of type $(26-d)$ can be conveniently stated at fixed 
$j$-cycle length $f_{j}(\sigma)$ in terms of exactly \emph {three
distinct classes} of $L$-cycles and their corresponding momenta. 
\\\\
\underline{Class I (exceptional)}.  For each cycle $L$ with 
\begin{equation}\lb{5.3}
\FLs = \tfrac{\fjs}{N},\;\, N=\textrm{integer}\geqslant 2
\end{equation}
we find the (exceptional) zero modes in cycle $j$
\begin{subequations}\lb{5.4}
\begin{gather}
\lb{5.4a}
\bigl\lbrace\jllj(0)\,,\;\hat{L}=0,1,\dots,\FLs-1\bigr\rbrace,\\[0.5em]
\lb{5.4b}
\dim\bigl\lbrace\jllj(0)\bigr\rbrace=\FLs.
\end{gather}
\end{subequations}
In this class, all indices $\hat{L}$ give zero modes, but only those particular values $\hj'=N\hat{L}$ which satisfy the constraints in Eq.~(5.2).
%manu5.4
\\\\
\underline{Class II (exceptional)}. 
For each $L$-cycle with 
\begin{equation}\lb{5.5}
\FLs = N\fjs,\;\, N=\textrm{integer}\geqslant 1
\end{equation}
we find the (exceptional) zero modes in cycle $j$
\begin{subequations}\lb{5.6}
\begin{gather}
\lb{5.6a}
\bigl\lbrace J_{\tfrac{\hj'\FLs}{\fjs},Lj}(0)\,,\;\hj'=0,1,\dots,\fjs-1\bigr\rbrace,\\[0.5em]
\lb{5.6b}
\dim\bigl\lbrace \hat{J}_{\tfrac{\hj'\FLs}{\fjs},Lj}(0)\bigr\rbrace=\fjs.
\end{gather}
\end{subequations}
\\\\
%manu5.5
\underline{Class III (generic)}.  In the generic $L$-cycle, neither $(F_L/f_j)$ nor $(f_j/F_L)$ is an integer, and we find that the only zero modes are the $\hat{L}=\hj'=0$ currents $J_{0Lj}(0)$ with
\begin{equation}\lb{5.7}
\dim \bigl\lbrace J_{0Lj}(0) \bigr \rbrace =1
\end{equation}
for each generic $L$-cycle.

We note in particular that every $L$-cycle, in any of the three classes, contains at least one zero mode.

In the general zero-mode classification of Ref.~[2], the generic class 
III  here provides examples of  general type I, while both the exceptional classes I 
and III here are examples of  general type II.

%manu5.6
The three classes  of $L$-cycles above can simply be described as follows:
%We can be more explicit about the lengths $\FLs$ of $L$-cycles which contribute in each class
\begin{subequations}\lb{5.8}
\begin{align}
\lb{5.8a}
(\textrm{Class I}):& \;\{\FLs=\textrm{the divisors of }\fjs \textrm{ except } \fjs \textrm{ itself}\}\\
\lb{5.8b}
(\textrm{Class II}):& \;\{\FLs=n\fjs,\, n \textrm{ a positive integer}\}\\
\lb{5.8c}
(\textrm{Class III}):& \;\{\textrm{all other }\FLs\}.
\end{align}
\end{subequations}
Note that $L$-cycles with length $\FLs=\fjs$ are counted in Class II. We illustrate with a few simple examples.
For $\fjs=2$, we find
\begin{subequations}\lb{5.9}
\begin{align}
\lb{5.9a}
(\textrm{Class I}):& \;\{\FLs=1\}\\
\lb{5.9b}
(\textrm{Class II}):& \;\{\FLs=\textrm{even}\geqslant 2\}\\
\lb{5.9c}
(\textrm{Class III}):& \;\{\FLs=\textrm{odd}\geqslant 3\}.
\end{align}
\end{subequations}
Similarly for any $\fjs$ prime we find 
\begin{subequations}\lb{5.10}
\begin{align}
\lb{5.10a}
(\textrm{Class I}):& \;\{\FLs=1\}\\
\lb{5.10b}
(\textrm{Class II}):& \;\{\FLs=N\fjs\,,\: N\;\textrm{a positive integer}\}\\
\lb{5.10c}
(\textrm{Class III}):& \;\{\textrm{all other }\FLs\}
\end{align}
\end{subequations}
and for $\fjs=8$:
\begin{subequations}\lb{5.11}
\begin{align}
\lb{5.11a}
(\textrm{Class I}):& \;\{\FLs=1,2,4\}\\
\lb{5.11b}
(\textrm{Class II}):& \;\{\FLs=8N\,,\: N\;\textrm{a positive integer}\}\\
\lb{5.11c}
(\textrm{Class III}):& \;\{\textrm{all other }\FLs\}.
\end{align}
\end{subequations}
Further examples are easily worked out by the reader.

%manu5.8
With this classification, we can write down an expression for the 
contribution of type $(26-d)$ to the target space-time dimension of cycle $j$ in sector $\sigma$:
\begin{subequations}\lb{5.12}
\begin{gather}
\lb{5.12a}
\hat{D}_j\ps^{(26-d)}=D_j\ps^{(26-d)}=\sum_{L(\textrm{I})}\FLs+\sum_{L(\textrm{II})}\fjs+\sum_{L(\textrm{III})}1\\[0.5em]
\lb{5.12b}
=\sum_{L(\textrm{I})}\FLs+\fjs N_{\textrm{II}}\ps^{'}+N_{\textrm{III}}\ps^{'},\\[0.7em]
\lb{5.12c}
N\ps^{'}=N_{\textrm{I}}\ps^{'}+N_{\textrm{II}}\ps^{'}+N_{\textrm{III}}\ps^{'},\\[0.7em]
\lb{5.12d}
D_j\ps^{(26-d)}\geqslant N \ps^{'}.
\end{gather}
\end{subequations}
Here the sums are over the $L$-cycles of Class I, II, or III, e.g. $\sum_{L(\textrm{I})}=N_{\textrm{I}}\ps^{'}$ and $N\ps^{'}$ is, as above,
the total number of $L$-cycles in the chosen element  $\omega\ps_{26-d}\in \Hp$.  The final lower bound in Eq.~(5.12d) follows
because every $L$-cycle has
at least one zero mode, and the equality is realized when all $L$-cycles in $\omega\ps_{26-d}$ are generic.

%manu5.9
Let us evaluate the result (5.12) in some simple cases, as a function of $j$-cycle length $\fjs$.  We begin with the trivial cycle length:
\begin{equation}\lb{5.13}
\fjs=1 :\quad D_j\ps^{(26-d)}=N\ps^{'}
\end{equation}
%where we remind that $N\ps^{'}$ is the total number of $L$-cycles in cycle $j$ of sector $\sigma$.
In this case, all the
contributions are from Class II, and the result in Eq.~(5.13) applies to the trivial element of any $H({\rm perm})_K$ with 
\begin{equation}\lb{5.14}
j=0,1,\dots,K-1\,,\; \bhj=0
\end{equation}
as well as many cycles of unit length in the elements of $H({\rm perm})_K=S_K$.

%manu5.9'
We consider next some examples of $j$-cycle length $\fjs$ as they occur in some full orbifold-string theories, for example
\begin{equation}\lb{5.15}
\left[\frac{{\rm U}(1)^{26K}}{H_+}\right]_{\rm{open}}\;, \quad
H_+\subset \Z_K \times (\pm\thickone)_d \times H({\rm perm})^{'}_{26-d}\,.
\end{equation}
Having already discussed the trivial element of $H({\rm perm})_K=\Z_K$, we focus now only on the $K-1$ non-trivial elements of $\Z_K$.  I 
%manu5.9''
begin with $K=\rm{prime}$, for which all $K-1$ non-trivial elements of $\Z_K$ have a single cycle $j=0$.  For each of these sectors we find: 
\begin{equation}\lb{5.16}
f_0\ps=K=\rm{prime} : \quad D_0\ps^{(26-d)}=N_{\{\FLs=1\}}\op+KN_{\{\FLs=nK\}}\op+N_{\rm{other}}\op\,.
\end{equation}
Note in particular that for the single non-trivial element of the simplest case  $H({\rm perm})_2=\Z_2$, this can be simplified to
\begin{equation}\lb{5.17}
f_0\ps=2: \quad D_0\ps^{(26-d)}=2N_E\ps^{'}+N_O\ps^{'},
\end{equation}
where $N_{E,O}\ps^{'}$ are respectively the number of $L$-cycles of even and odd length $\FLs$ in $\omega\ps\in H({\rm perm})_{26-d}^{'}$.  This
result is in agreement with the counting of zero modes in Ref.~[5].

%manu5.10
For $H({\rm perm})_4=\Z_4$, we have 2 non-trivial single $j$-cycle elements with 
\begin{equation}\lb{5.18}
f_0\ps=4: \quad D_0\ps^{(26-d)}=N_{\{\FLs=1\}}\op+2N_{\{\FLs=2\}}\op+4N_{\{\FLs=4n\}}\op+N_{\rm{other}}\op
\end{equation}
and one sector with 2 $j$-cycles of length 2, each of which is described by Eq.~(5.17)  

For $H({\rm perm})_8=\Z_8$, we have one non-trivial sector with a single $j$-cycle of length $f_0(\sigma)=8$ and hence
\begin{equation}\lb{5.19}
D_0\ps^{(26-d)}=N_{\{1\}}\op+2N_{\{2\}}\op+4N_{\{4\}}\op+8N_{\{8n\}}\op+N_{\rm{other}}\op
\end{equation}
where $N_{\{1\},\{2\},\{4\},\{8n\}}\op$ are respectively the number of $L$-cycles of length 1, 2, 4, and multiples of 8.
The other non-trivial elements of $\Z_8$ include 5 two-cycle elements ($j=0,1$), each of which has $\fjs=4$ (use Eq.~(5.18)) 
and one element with 4 $j$-cycles of length $f_j=2$ (use Eq.~(5.17)).  In this way, the non-trivial sectors of $\Z_K,K=2^n$ are easily worked out
%manu5.11
by induction from the results of $K=2^{n-1}$, adding at each step only the single $j$-cycle result for $f_{0}(\sigma)$=$2^n$.

%manu5.12
We may also write down the corresponding contribution of type $(26-d)$ to the
momentum-squared operator [2] of cycle $j$ in sector $\sigma$:
\begin{subequations}\lb{5.20}
\begin{gather}
\lb{5.20a}
\hskip-13em{}\hat{P}^2_j\ps^{(26-d)}=P^2_j\ps^{(26-d)}\\[0.5em]
\lb{5.20b}
\begin{gathered}[b]
\hskip-10em{}=-\Bigl\lbrace \sum_{L(\rm{I})} \sum_{\hat{L}=0}^{\FLs-1}\tilde{J}_{\hat{L}Lj}(0) \,\tilde{J}_{-\hat{L},Lj}(0)\\[0.7em]
+\sum_{L(\rm{II})} \sum_{\hj'=0}^{\fjs-1}
 \tilde{J}_{\hj' \tfrac{\FLs}{\fjs},Lj}(0) \,\tilde{J}_{-\hj' \tfrac{\FLs}{\fjs},Lj}(0)\\[0.7em]
+ \sum_{L(\rm{III})}
 \tilde{J}_{0Lj}(0) \,\tilde{J}_{0Lj}(0)\Bigr\rbrace,
\end{gathered}\\[0.7em]
\lb{5.20c}
\tilde{J}_{\hat{L}Lj}(0)\equiv\tfrac{1}{\sqrt{\FLs}}J_{\hat{L}Lj}(0)\,.
\end{gather}
\end{subequations}
Here I have chosen for simplicity to rescale each of the zero modes  according to Eq.\eqref{5.20c}.

%manu5.13
Finally, the adjoint operations in Eq.~(2.15) imply
\begin{subequations}\lb{5.21}
\begin{gather}
\lb{5.21a}
\hskip-13em{}P^2_j\ps^{(26-d)\dagger}=P^2_j\ps^{(26-d)}\\[0.5em]
\lb{5.21b}
\begin{gathered}[b]
0\leqslant -P^2_j\ps^{(26-d)}=\sum_{L(\rm{I})} \sum_{\hat{L}=0}^{\FLs-1} 
\babs{J_{\hat{L}Lj}(0)}^2 +\\[0.7em]
\hspace{1in}+\sum_{L(\rm{II})} \sum_{\hj'=0}^{\fjs-1}
\babs{ \tilde{J}_{\hj' \tfrac{\FLs}{\fjs},Lj}(0)}^2 +
 \sum_{L(\rm{III})} \tilde{J}^2_{0Lj}(0)
\end{gathered}
\end{gather}
\end{subequations}
where the form (5.21b) of the momentum-squared holds on any state with diagonalized momenta.  
As anticipated above, the contribution of type $(26-d)$ to the momentum-squared operator is indeed space-like.

\section{The Integer-Moded Sequences}\lb{Section 6}
 We have seen in our large example that the target space-time of cycle 
 $j$ in sector $\sigma$ has the \emph{target space-time dimension}
\begin{equation}\lb{6.1}
\hat{D}_j\ps=D_j\ps=
\begin{cases}
d+D_j\ps^{(26-d)} \quad {\rm for} \quad \eps=0\quad or \quad \eps=1 \,\,{\rm and} \; \fjs \;{\rm even}\\[0.5em]
D_j\ps^{(26-d)} \quad {\rm for} \quad \eps=1 \;\, {\rm and} \;\fjs \;{\rm odd}
\end{cases}
\end{equation}
where general formulae for $D_j\ps^{(26-d)}$, $1\leqslant d \leqslant25$ are given in
the previous section, and $D_j\ps^{(0)}=0$ for the extremal case $d=26$.  One
correctly expects that the target space-times described in the first and 
second parts of Eq.~(6.1) are respectively
Lorentzian and Euclidean, which shall explore in some detail below.

%manu6.2
We know that the space-time dimensions in Eq.~(6.1) correspond to the zero modes of the orbifold-string systems, but
in the discussion which follows it will also be important to consider the natural extension of the zero modes to
the set of \emph{integer-moded sequences} $\{J_j(M)_\sigma\}$ of cycle 
$j$ in sector $\sigma$, which include the zero modes  $\{J_j(0)_\sigma\}$ 
when $M=0$.
%which are in 1-1 correspondence
%with the zero modes $\{J_j(0)_\sigma\}$ of cycle $j$ in sector $\sigma$. 
The final factors in the twisted current
algebras (2.14b,d) show immediately that all the integer-moded sequences commute with all the fractional (non-integer) moded sequences
\begin{equation}\lb{6.2}
\bigl[\{J_j(M)_\sigma\}, \{J_\ell(\textrm{fractional-moded})_\sigma\}\bigr]=0\;,\;\; \forall j,\ell,\sigma.
\end{equation}
The current algebra (2.14b) also tells us directly that the integer-moded sequences of type~$\!(d)$ are ordinary
untwisted string coordinates, and moreover, we find after some algebra that the integer-moded
%manu 6.3
sequences of type $\!(26-d)$ can be rescaled and relabeled as a set of $D_j\ps^{(26-d)}$ \emph{additional} ordinary untwisted
string coordinates:
\begin{equation}\lb{6.3}
\{J_{aj}(M)\equiv \tfrac{1}{\sqrt{F_J\ps}}J_{\hat{J}Jj}(M)\,,\; a=d,\dots,d+D_j\ps^{(26-d)}-1\}.
\end{equation}
The rescaled integer-moded sequences (6.3) include the rescaled zero-modes of type $\!(26-d)$ in Eq.~(5.20c).

Then our large example can be broken into two distinct physical cases, as discussed in the following two sections.

%manu7.1
\section{Lorentzian Space-Time Symmetry $SO(D_j\ps-1,1)$}

Following the previous discussion, this section is limited to the discussion of the following two cases
\begin{equation}\lb{7.1}
\begin{split}
1)\; \eps=&0,\,\,\forall f_{j}(\sigma)\\[0.5em]
2)\; \eps=&1,\; \fjs\; \rm{even}
\end{split}
\end{equation}
for which we shall see that the dynamics of cycle $j$ is \emph{Lorentzian}.  We remind the reader
that the values $\eps=0,1$ correspond  respectively to $(\omega)_d=(\thickone)_d$ and $(-\thickone)_d$.

In these two cases, our large example can be put into the following form:
%manu7.2
\begin{subequations}\lb{7.2}
\begin{gather}
\lb{7.2a}
\begin{gathered}[b]
L_j(M)=-\tfrac{1}{2} \eta_{(D_j\ps)}^{ab}\sum_{Q\in\Z}\nosub{J_{a j}(Q) 
J_{b j}(M-Q)}{M}\,+\\[0.5em]
+\tfrac{1}{2}\Bigl(\sum_{L,\hat{L}}\Bigr)'\tfrac{1}{F_L\ps} \sum_{Q\in\Z}
\nosub{J_{\hat{L}Lj}(Q+\fjs\tfrac{\hat{L}}{\FLs}) 
\hat{J}_{-\hat{L},Lj}\bigl(M-Q-\fjs\tfrac{\hat{L}}{\FLs}\bigr)}{M}\,+\\[0.5em]
+\,\kd{M}{0}\,\hat{\delta}_{0j}\ps^{(26-d)}
\end{gathered}\\[0.7em]
\lb{7.2b}
\sum_L \FLs=26-d,\,\; D_j\ps=d+D_j\ps^{(26-d)},\,\; \eta_{(D_j\ps)}=
\Bigl(\hskip-2pt \begin{array}{rc}1 & 0 \\  0 & -\thickone \end{array} \Bigr)_{D_j\ps}.
\end{gather}
\end{subequations}
%manu7.2'
Here $D_j\ps$ is the \emph{target space-time dimension} of cycle $j$ in sector $\sigma$, $ \eta_{(D_j\ps)}$ is the
corresponding $D_j\ps$-dimensional flat \emph {Minkowski-space metric}, 
and the conformal-weight shifts of type $(26-d)$ are given in Eq.~(2.7d).  The generators $\{L_j(M)\}$ are
Virasoro with uniform cycle central charge $\cjs=26$
\begin{equation}\lb{7.3}
\bigl[ L_j(M),L_\ell (N) \bigr]=\delta_{j \ell}\bigl\lbrace(M-N)L_j(M+N)+\tfrac{26}{12}M(M^2-1)\,\kd{M+N}{\,0}\bigr\rbrace
\end{equation}
and the two kinds of currents satisfy the algebra:
%manu7.3
\begin{subequations}\lb{7.4}
\begin{gather}\lb{7.4a}
\bigl[ L_j(M),J_{a \ell} (N) \bigr]=-\delta_{j \ell} \,N J_{a \ell}(M+N),\quad a=0,1,\dots,\djs-1,\\[0.5em]
\lb{7.4b}
\bigl[ L_j(M),J_{\hat{L}L \ell} (N+\fls\hatLovfs) \bigr]=-\delta_{j \ell} (N+\fjs\hatLovfs)\, J_{\hat{L}L \ell} (M+N+\fjs\hatLovfs),\\[0.5em]
\lb{7.4c}
\bigl[ J_{aj}(M),J_{b \ell} (N) \bigr]=\delta_{j \ell}\,N\, \eta_{ab}^{(\djs)}\kd{M+N}{\,0},\\[0.5em]
\lb{7.4d}
\begin{gathered}[b]
\hskip-15em{}\bigl[ \jjjj(M+\fjs\hatJovfs),J_{\hat{L}L \ell} (N+\fls\hatLovfs) \bigr]=\\[0.5em]
\delta_{j\ell }\,\delta_{JL}\, F_J\ps\bigl(M +\fjs\hatJovfs \bigr)\,
\kd{\hat{J}+\hat{L}}{\,0\bmod F_J\ps \phantom{\tfrac{1}{2}}}\!\!\kd{M+N+\fjs\frac{\hat{J}+\hat{L}}{F_J\ps}}{\,0}\;.
\end{gathered}
\end{gather}
\end{subequations}
%manu7.3'
The $\djs$-dimensional Minkowski metric is seen again in Eq.~(7.4c).

%manu7.4
In this system, we have collected all the integer-moded sequences ($d$ of type $\!(d)$ plus $\djs^{(26-d)}$ of type $\!(26-d)$) 
and used the rescaling/relabeling (6.3) to write all the integer-moded sequences in a unified notation as
$\djs$ \emph{ordinary untwisted string coordinates}.  See in particular 
the ordinary untwisted current algebra of all the
integer-moded sequences in Eq.~(7.4c) and the \emph{first term} of the Virasoro generators (7.2a), which can be put in
this ``ordinary form'' after a shift under the sum on $Q$. ( The shift, by the integer $(\fjs/2),\,\fjs$ even, is 
only necessary for $\epsilon=1$.)  We remind the reader
that this ``ordinary'' term does not contribute to the conformal-weight shift $\hat{\delta}_{0j}\ps=\hat{\delta}_{0j}\ps^{(26-d)}$ in the third term
of the Virasoro generators -- which comes entirely from the 
fractional-moded sequences. Similarly, the ground-state mass-squared (1.12c) 
of cycle $j$ in sector $\sigma$
\begin{equation}
\lb{7.5}
P_j^{2}(\sigma)_{(0)} = 2(-1+\hat{\delta}_{0j}\ps^{(26-d)}) \geq-2
\end{equation}
is shifted only by the fractionally-moded sequences.

%manu7.5
The prime on the sum in the second term of the Virasoro generators (7.2a) denotes omission of the integer-moded sequences
of type $\!(26-d)$, which we understand are now included in the first 
(ordinary) term. The second term therefore sums only over the 
fractional-moded sequences of type $(26-d)$, whose twisted 
current algebra is given in Eq.~(7.4d).
%can now be read 
%only for the fractional-moded sequences. 
%The algebra of the fractional-moded currents
%is given in Eq.~(7.4d).  
Since the integer-moded sequences commute with the fractional-moded sequences, we 
see that the first (ordinary) term of the Virasoro generators,
as well as the combination of the second and third terms of the generators, form two commuting sets of Virasoro
generators -- with central charges $\djs$ and $(26-\djs)$ respectively.

%manu7.6
It is clear that the first term in Eq.~(7.2a) is a set of \emph {ordinary untwisted Virasoro generators on a Lorentzian target space of
dimension $\djs$}.  Therefore the dynamics of cycle $j$ in sector $\sigma$ has an $SO(D_j\ps-1,1)$ \emph{Lorentz symmetry}, under which
the integer-moded currents $\{J_{aj}(M)\}$ transform as  Lorentz-vectors of dimension $\djs$, while
the fractional-moded currents transform as \emph{scalars} under the 
Lorentz group. This conclusion is one of the central results of this paper.

Of course, in these cases the integer-moded sequences of type $\!(d)$ 
already possessed a ``primordial'' Lorentz symmetry $SO(d-1,1)$ so that the
true target space-time symmetry $SO(D_j\ps-1,1)$ of cycle $j$ in sector $\sigma$
can be understood as an \emph{enhanced Lorentz symmetry}.  This enhanced 
Lorentz symmetry is transparent in the system only after
we have collected (and rescaled) the integer-moded sequences both of type 
$\!(d)$ and type $\!(26-d)$. We emphasize that the enhanced 
symmetry is a completely natural one, given 
the total number $\djs$ of target space-time dimensions. See Ref.~[2] for an
outline of the target 
space-time dimensions in more general orbifold-string 
theories of permutation-type.

%manu7.6'
Here we have discussed the enhanced Lorentz symmetry of cycle $j$ only in the equivalent, reduced formulation of the cycle
at $\cjs=26$.  Using the inverse of the map (1.5), it is also possible to see this symmetry in the original formulation
of the cycle at $\hcjs=26\fjs$ -- a subject to which we will return elsewhere.

%manu7.6''
With their ordinary $\djs$-dimensional string subspaces and $26-\djs$ extra
twisted scalar currents, these Lorentzian systems may appear at first sight to 
resemble some ordinary target-space string compactification which leaves only $\djs$ continuous dimensions.  But in fact
the sector structure and characteristic modeing 
$(f_{j}(\sigma)\hat{L})/F_{L}(\sigma)$ of the extra scalar currents shows 
clearly that these systems are quite distinct from compactification, and 
generically new for all $H({\rm perm})_{K}$ and $H({\rm perm})'_{26-d}$.
%these systems, which generalize our previous results [5] for $H({\rm 
%perm})_2=\Z_2$, are generically \emph{new} and quite distinct from compactification.
This conclusion generalizes our earlier statement [5] for the simple 
cases $H({\rm perm})_{2}=\Z_2$ or $\Z_{2}({\rm w.s.})$.
 
%manu7.7
I close this section with a few remarks about negative-norm states in 
these Lorentzian string systems. 

For the Lorentzian cycles, the adjoint operations include the generalized 
hermiticity (2.15c) of the reduced Virasoro generators as well as
\begin{subequations}\lb{7.6} 
\begin{gather}\lb{7.6a}
J_{aj}(M)^\dagger=J_{aj}(-M),\quad a=0,1,\dots,\djs-1\\[0.5em]
\lb{7.6b}
\jllj(M+\fjs\hatLovfs)^\dagger=J_{-\hat{L},Lj}(-M-\fjs\hatLovfs)
\end{gather}
\end{subequations}
and the periodicity condition (2.15b) continues to hold for the fractional sequences.  Then, following our
discussion in Sec.~2, we see that the negative-norm basis states occur only when the basis state
contains an odd number of "ordinary" time-like modes $\{J_{0j}(M<0)\}$.  If $\djs\leqslant 26$, which we have
\emph{not} established here beyond the "pure" permutation orbifolds [2] with $\djs=26$, then one may continue
to expect [3] that all negative-norm physical states will decouple.  We shall continue this discussion in the following
paper, showing in fact that $\djs\leqslant 26$ for the all the open-and 
closed-string  sectors in the large example of 
the orientation-orbifold string systems [5] -- i.e. $H({\rm 
perm})_2=\Z_2(w.s.)$ and all $H'_{26-d}$.

%manu8.1
\section{Euclidean Space-Time Symmetry $SO(\djs^{(26-d)})$}\lb{Section 8}

Beyond the Lorentzian cases in Eq.~(7.1), there remains to study the 
cycles $j$ of sector $\sigma$ in the case
%For the cycles $j$ in sector $\sigma$ with 
 \begin{equation}\lb{8.1}
\eps=1,\; \fjs\; \rm{odd}
\end{equation}
which, as we shall see, leads to more exotic strings.

In this case we find the following algebraic systems,
beginning with the reduced Virasoro generators at cycle central charge $c_j(\sigma)=26$:
%manu8.2
\begin{subequations}\lb{8.2}
\begin{gather}
\lb{8.2a}
\begin{gathered}[b]
L_j(M)=\tfrac{1}{2} \delta^{ab}_{(\djs)} \sum_{Q\in\Z}\nosub{J_{a j}(Q) 
J_{b j}(M-Q)}{M} +\\[0.5em]
-\tfrac{1}{2} \eta_{(d)}^{ab}\sum_{Q\in\Z}\nosub{J_{1 a 
j}(Q+\tfrac{\fjs}{2})J_{-1,b j}(M-Q-\tfrac{\fjs}{2})}{M} +\\
+\tfrac{1}{2}\Bigl(\sum_{L,\hat{L}}\Bigr)'\tfrac{1}{F_L\ps} \sum_{Q\in\Z}
\nosub{J_{\hat{L}Lj}(Q+\fjs\tfrac{\hat{L}}{\FLs})
\hat{J}_{-\hat{L},Lj}\bigl(M-Q-\fjs\tfrac{\hat{L}}{\FLs}\bigr)}{M} +,\\[0.5em]
+\,\kd{M}{0}\,(\tfrac{d}{16}+\hat{\delta}_{0j}\ps^{(26-d)}),
\end{gathered}\\[0.7em]
\lb{8.2b}
\sum_L \FLs=26-d,\;\,\, D_j\ps=D_j\ps^{(26-d)},\;\,\, \delta_{(\djs)}=(\thickone)_{\djs}.
\end{gather}
\end{subequations}
%manu8.3
We remind that the contributions of type $(26-d)$ to the conformal-weight 
shifts and the target-space dimensionalities are given respectively in 
Eqs.~(2.7d) and (5.12).
The non-zero commutators of these modes are as follows:
%manu8.4
\begin{subequations}\lb{8.3}
\begin{gather}
\lb{8.3a}
\bigl[ L_j(M),J_{a \ell} (N) \bigr]=-\delta_{j \ell} \,N J_{a j}(M+N),\\[0.5em]
\lb{8.3b}
a=d,d+1,\dots,d+\djs^{(26-d)}-1,\\[0.5em]
\lb{8.3c}
\bigl[ L_j(M),J_{1 a \ell} (N+\tfrac{\fls}{2}) \bigr]=-\delta_{j \ell} (N+\tfrac{\fjs}{2}) J_{1 a j}(M+N+\tfrac{\fjs}{2}),\\[0.5em]
\lb{8.3d}
a=0,1,\dots,d-1,\\[0.5em]
\lb{8.3e}
\bigl[ L_j(M),J_{\hat{L}L \ell} (N+\fls\hatLovfs) \bigr]=-\delta_{j \ell} (N+\fjs\hatLovfs)\, J_{\hat{L}L \ell} (M+N+\fjs\hatLovfs),\\[0.5em]
\lb{8.3f}
\bigl[ J_{aj}(M),J_{b \ell} (N) \bigr]=\delta_{j \ell}\,M \, \delta_{ab}^{(\djs)},\\[0.5em]
\lb{8.3g}
\bigl[ J_{1aj}(M+\tfrac{\fjs}{2}),J_{1 b \ell} (N+\tfrac{\fls}{2})\bigr]=
-\delta_{j\ell }\,\eta_{ab}^{(d)} (M+\tfrac{\fjs}{2})\kd{M+N+\fjs}{\,0}\\[0.5em]
\lb{8.3h}
\begin{gathered}[b]
\hskip-15em{}\bigl[ \jjjj(M+\fjs\hatJovfs),J_{\hat{L}L \ell} (N+\fls\hatLovfs) \bigr]=\\[0.5em]
\delta_{j\ell }\,\delta_{JL}\, F_J\ps\bigl(M +\fjs\hatJovfs \bigr)\,
\kd{\hat{J}+\hat{L}}{\,0\bmod F_J\ps \phantom{\tfrac{1}{2}}}\kd{M+N+\fjs\frac{\hat{J}+\hat{L}}{F_J\ps}}{\,0}\;.
\end{gathered}
\end{gather}
\end{subequations}
%manu8.5
We supplement the mode algebra with the adjoint operations
\begin{subequations}\lb{8.4}
\begin{gather}
\lb{8.4a}
J_{aj}(M)^\dagger=J_{aj}(-M), \quad a=d,d+1,\dots,d+\djs-1,\\[0.5em]
\lb{8.4b}
J_{1aj}(M+\tfrac{\fjs}{2})^\dagger=J_{-1,aj}(-M-\tfrac{\fjs}{2}), \quad a=0,1,\dots,d-1,\\[0.5em]
\lb{8.4c}
\jllj(M+\fjs\hatLovfs)^\dagger=J_{-\hat{L},Lj}(-M-\fjs\hatLovfs)
\end{gather}
\end{subequations}
which follow, along with the generalized hermiticity (2.15c) of the 
Virasoro generators, from the adjoint operations in Eqs.~(2.15d,e).  Similarly, the periodicity conditions for these modes are easily worked out from
the corresponding relations in Eqs.~(2.15a,b).

%manu8.6
These orbifold-string systems are quite different from those of the 
previous section, first because they do \emph{not} have an ordinary Lorentzian
string subsector.

%manu8.7
In further detail, these cycles have $\djs^{(d)}=0$, i.e. there are no 
integer-moded sequences of type $(d)$.  Therefore all integer-moded sequences
are of type $\!(26-d)$, and these target space-times are \emph{Euclidean} with Euclidean space-time dimension $\djs=\djs^{(26-d)}$.
After the relabeling (6.3), the integer-moded sequences
$\{J_{aj}(M),\;\, a=d,\dots,d+\djs^{(d)}-1\}$ of type $\!(26-d)$ are collected in the first term of the Virasoro
generators (8.2a), and their Euclidean current algebra is given in 
Eq.~(8.3f). The Euclidean space-time metric $\delta_{(D_j(\sigma))}$ 
(Kronecker delta) is seen in both places. The half-integer-moded
currents $\{J_{1aj}(M+\tfrac{\fjs}{2}),\;\, \fjs\, \rm{odd}\}$ appear in the second term of Eq.~(8.2a) as the entire
contribution of type $\!(d)$.  Again, the prime on the sum in the third term of Eq.~(8.2) denotes the omission
of the integer-moded sequences of type $\!(26-d)$, which we understand are now included in the first term of the Virasoro generators.

%manu8.8
These Euclidean cycles have a \emph{Euclidean target space-time symmetry} $SO(\djs^{(26-d)})$, under
which the integer-moded sequences $\{J_{aj}(M),\;\, 
a=d,\dots,d+\djs^{(26-d)} -1\}$ transform as $\djs^{(26-d)}$-dimensional
vectors. All remaining modes, including both the half-integer moded sequences $\{J_{1aj}(M+\tfrac{\fjs}{2}), a=0,1,\dots,d-1\}$ and
the fractional-moded sequences $\{\jllj \bigl(M+(\fjs\hat{L}/\FLs 
\bigr)\}$, are scalars under this orthogonal group. 

%manu8.9
The Virasoro generators (8.2) are in fact a sum of three sets of commuting Virasoro generators.  The first (Euclidean) term
of Eq.~(8.2a) is by itself Virasoro with central charge
\begin{equation}\lb{8.5}
c_{j1}\ps=\djs^{(26-d)}.
\end{equation}
The second (half-integer moded) term of Eq.~(8.2a) together with the $(d/16)$ contribution to the 
conformal-weight shift form a second set of Virasoro generators with central charge
\begin{equation}\lb{8.6}
c_{j 2}\ps=d.
\end{equation}
Finally, a third set of Virasoro generators with 
\begin{equation}\lb{8.7}
c_{j3}\ps=26-d-\djs^{(26-d)}
\end{equation}
is formed from the third term of Eq.~(8.2a) and the contribution $\hat{\delta}_{0j}\ps^{(26-d)}$ to the conformal-weight shift.  It
is easily checked then that 
\begin{equation}\lb{8.8}
c_{j1}\ps+c_{j2}\ps+c_{j3}\ps=\cjs=26
\end{equation}
as required for the reduced description of cycle $j$ in sector $\sigma$.

%manu8.10
The $d$-dimensional Lorentz metric $\eta_{(d)}$ appears in the second term of the Virasoro generators and again in the current algebra (8.3g).  This
means that basis states formed with an odd number of time-like scalar operators $\{J_{1 0 \ell}((M+\tfrac{\fls}{2})<0) \}$ have negative norm.
As a consequence, in distinction to the Lorentzian cycles, we cannot 
expect all Euclidean cycles to be free of negative-norm physical states. 
We will give an example of this circumstance below.

%manu8.11
This brings us to another salient feature of these Euclidean cycles, namely that
they consist of at most \emph{a finite number of physical states}, and 
these states are mostly tachyonic because the Euclidean momenta satisfy $P^2_j\ps\leqslant 0$. 

%manu8.12
To see this, we begin with the zeroth component of the physical-state condition (1.6a) in the schematic form
\begin{subequations}\lb{8.9}
\begin{gather}
\lb{8.9a}
L_j(0)=\tfrac{1}{2}\bigl(-P^2_j\ps+R_{j}\ps \bigr)+\tfrac{d}{16}+\hat{\delta}_{0j}\ps^{(26-d)}=1\\[0.5em]
\lb{8.9b}
-2 \leqslant P^2_j\ps \leqslant 0, \quad \hat{\delta}_{0j}\ps^{(26-d)} \geqslant 0, \quad  R_{j}\ps \geqslant 0.
\end{gather}
\end{subequations}
%manu8.13
Here we are assuming that the physical state is diagonal in the momenta 
and the generalized number 
operator $R_{j}\ps$, so that we may treat these quantities as numbers. 
The lower bound given for the momentum-squared is the lower bound on the 
ground-state momentum-squared in Eq.~(1.12c), while the upper bound is 
the Euclidean character of the momenta. 
%manu8.14
That these cycles have at most a finite number of physical states is then 
clear from the strict positivity of the increments (level-spacing) $\Delta(P^2_j\ps)$ in Eq.~(2.17).

This can equivalently be seen from the following closely-related double 
inequality on the number operator
%the double inequality on the number operator
\begin{equation}\lb{8.10}
0 \leqslant R\ps \leqslant 2 \bigl(1-\tfrac{d}{16}-\hat{\delta}_{0j}\ps^{(26-d)}\bigr)
\end{equation}
and the fact that the increments $\Delta(R_{j}(\sigma))=\Delta(P^2_j\ps)$ in Eq.~(2.17c) are strictly positive.  The Lorentzian
cycles of the previous section have no such upper bound on the momenta or 
the number operator, both of which follow
immediately from the Euclidean character of the momenta.

%manu8.15
According to Eq.~(8.10), there will be in fact no physical states at all in a given Euclidean cycle $j$ unless the conformal-weight shifts
and parameter $d$ satisfy
\begin{subequations}\lb{8.11}
\begin{gather}
\lb{8.11a}
0 \leqslant \hat{\delta}_{0j}\ps^{(26-d)} \leqslant 1-\tfrac{d}{16},\\[0.5em]
\lb{8.11b}
1 \leqslant d \leqslant 16 \\[0.5em]
\lb{8.11c}
10 \leqslant (26-d) \leqslant 25
\end{gather}
\end{subequations}
where Eqs.~(8.11b,c) follow from Eq.~(8.11a).
The upper bound in Eq.~(8.11a) is non-trivial, and we remind that the explicit form of $\hat{\delta}_{0j}\ps^{(26-d)}$ is 
given in Eq.~(2.7d).  
We should also emphasize what is implicit in the discussion above, namely that these 
conditions are sufficient to 
guarantee the existence of an oscillator-free tachyonic
physical ground state with ground-state momentum-squared:
\begin{subequations}\lb{8.12}
\begin{gather}
\lb{8.12a}
\kets{\chi\ps}{j}=\ket{0,J_j(0)_\sigma}\\[0.5em]
\lb{8.12b}
-2 \leqslant P^2_j\ps_{(0)}=2\bigl(-1+\tfrac{d}{16}+\hdz \bigr) \leqslant 0.
\end{gather}
\end{subequations}
%as given earlier in Refs. [5,2] and Eq.~(1.12c).
%The Euclidean upper bound here is equivalent to the upper bound in 
%Eq.~(8.11a), and this statement is in accord with 
The first parts of this statement are in accord with
Eqs.~(1.12c),(2.7b) and (4.5b),
while the Euclidean upper bound is equivalent to the upper bound in Eq.~(8.11a). 

%manu8.15'
In summary, the Euclidean cycles with
\begin{equation}\lb{8.13}
\eps=1,\quad \fjs\;{\rm odd},\quad 16 \geqslant d \geqslant 1, \quad 25 \geqslant (26-d) \geqslant 10
\end{equation}
consist of at most a short collection of tachyonic states with the 
highest possible states being some number of zero-mass particles.
It is clear that many non-trivial short Euclidean cycles of this type exist 
but, as noted above, they may not all be free of negative-norm physical states. 
For example choosing $d=f_{j}(\sigma)=1$ and the unit 
element of $H({\rm perm})_{25}'$ with all $F_{L}(\sigma)=1$, 
we find that $\hat{\delta}_{0j}(\sigma)^{(25)}=0 \leq 15/16$ and the
tachyonic ground state is found at $ P^2_j\ps_{(0)}=-15/8$. The rest of 
this short collection consists of  
a single excited tachyonic physical state $J_{10j}(-1/2)|P^{2}_{j}(\sigma)=-7/8>$ with 
negative norm. 

The particular example above can be eliminated by judicious choice of divisors for the full 
orbifold, but non-trivial short Euclidean cycles deserve further 
study elsewhere -- not least because 
they apparently represent a crossover point between string theory and particle theory.   
A related question of interest is the fate of such cycles in the 
as-yet-unconstructed orbifold-superstring theories of permutation-type [3], where one may 
suspect that the short collections consist only of zero-mass particles.   

%manu8.16
We finally note that an even more limited spectrum is 
possible for certain Euclidean cycles: The conditions in Eqs.~(8.11b,c) tell us immediately that
there are \emph{no physical states at all} for any Euclidean cycle with 
\begin{equation}\lb{8.14}
\eps=1,\quad \fjs\;{\rm odd},\quad 26 \geqslant d \geqslant 17, \quad 9 \geqslant (26-d) \geqslant 0.
\end{equation}
In these cases, we refer to the Euclidean cycle as an \emph{extinguished 
cycle}, and many examples of such cycles are easily constructed from the 
conditions (8.14). The following section discusses a simple class of these
extinguished cycles in some detail.

\section{Null Cycles}\lb{Section 9}

Among the extinguished Euclidean cycles (with no physical states) in 
Eq.~(8.14), we focus here on a simple set of subexamples
with \emph{no momenta at all} $(P_j\ps=0)$
\begin{equation}\lb{9.1}
\djs=\djs^{(d)}=\djs^{(26-d)}=\{J_j(0)_\sigma\}=P^2_j\ps=0
\end{equation}
which we call \emph{null cycles}.   Such null cycles are rare among the 
extinguished Euclidean cycles, but all the null 
cycles in our large example are easily
located with the zero-mode counting above:
\begin{equation}\lb{9.2}
\djs=0: \quad \eps=1, \quad \fjs\; {\rm odd},  \quad H({\rm perm})^{'}_0 \;\; {\rm trivial},  \quad d=26.
\end{equation}
In particular, we have used the fact that every $L$-cycle has at least 
one zero mode (see Sec.~5) to obtain the
requirement that for null cycles $H({\rm perm})_{26-d}^{'}$ must be trivial and hence $d=26$.

%manu9.2
It is instructive to consider the fate of some full orbifolds which contain at least some of these extinguished null cycles, for
example the generalized permutation orbifolds
\begin{equation}\lb{9.3}
\frac{{\rm U}(1)^{26 K}}{H_+}, \quad H_+=\bigl\lbrace \{ \omega\ps\in\Z_K \} \times (\thickone)_{26}; \; \{ \omega\ps\in\Z_K \} \times (-\thickone)_{26} \bigr\rbrace, \quad K \; {\rm odd}
\end{equation}
%manu9.3
where we have chosen $H({\rm perm})_K=\Z_K$ for $K$ odd and displayed all the elements of the divisor $H_+$.  In these examples, all the cycles of all
the sectors after the semi-colon are null extinguished cycles, with no 
momenta and no physical particles.  This follows because all the elements
of the odd cyclic groups
%$\Z_K$, $K$ odd
have cycle lengths $\fjs$ odd, including the trivial element with $\fjs=1$.  In this latter case, the
sector arises from the action of $(\omega)_{26}=(-\thickone)_{26}$ on all 26 dimensions of the original closed string $\uts$, and
the only currents are half-integral moded scalar fields
$\{J_{1 a j}(M+\tfrac{1}{2}), \; a=0,\dots,25, \; j=0,\dots,K-1 \}$ -- a 
familiar case [23] with no zero modes.

%manu9.4
Thus, \emph{as string systems}, the set of generalized cyclic 
permutation-orbifolds (9.3) degenerate to the "pure" cyclic
permutation-orbifolds
\begin{equation}\lb{9.4}
\frac{U(1)^{26 K}}{\Z_K}, \quad K \; {\rm odd}
\end{equation}
each cycle of which is in fact known [7,2] to be equivalent to an ordinary 26-dimensional closed string. 
%manu9.5

I mention in conclusion the intuitive parallel of null cycles to \emph{spurions} at $\{ P_\mu=0 \}$.

\section{Conclusions and Directions}

In both the unreduced and the equivalent, reduced formulations we have provided the physical-state conditions and Virasoro
generators of the following large examples of bosonic orbifold-string 
systems of permutation type [3,2]:
\begin{subequations}\lb{10.1}
\begin{gather}
\lb{10.1a}
\frac{U(1)^{26 K}}{H_+}, \quad \left[\frac{{\rm U}(1)^{26K}}{H_+}\right]_{{\rm open}}\\[0.5em]
\lb{10.1b}
H_+ \subset \bigl\lbrace H({\rm perm})_K \times (\thickone)_d \times H({\rm perm})_{26-d}^{'}; \,
H({\rm perm})_K \times (-\thickone)_d \times H({\rm perm})_{26-d}^{'} \bigr\rbrace.
\end{gather}
\end{subequations}
Here the divisor $H_+$ can be any subgroup of the elements shown, and the complete orbifold-string
system has a sector $\sigma$ for each equivalence class of $H_+$.  The theories  
%$\left[\frac{{\rm U}(1)^{26K}}{H_+}\right]_{{\rm open}}$ are
$\left[U(1)^{26K}/H_{+}\right]_{{\rm open}}$ 
are open-string analogues [22,3] of the closed-string systems
% $\frac{U(1)^{26 K}}{\Z_K}$,
$U(1)^{26K}/H_{+}$, the latter requiring
the addition of right-mover copies of the mode systems given above. Our 
analysis also includes the orientation-orbifold string
systems with $H_{+}\rightarrow H_{-}$ and $H({\rm perm})_{K}\rightarrow \Z_{2}({\rm w.s.})$,
on which we comment separately below.

The present paper is an extension of our prior construction [5] at $K=2$, now 
emphasizing the target space-times and target space-time symmetries of the theories (10.1) for all $K$. In the discussion above, the elements
of both permutation groups $H({\rm perm})_K$ and $H({\rm perm})_{26-d}^{'}$ are analyzed
in terms of their respective cycle lengths
\begin{equation}\lb{10.2}
\sum_j \fjs=K, \quad \sum_L \FLs=26-d, \quad \fjs \geqslant 1,  \quad \FLs \geqslant 1
\end{equation}
and we remind that the elements $(\omega)_d=(\thickone)_d$ and $(-\thickone)_d$
%manu10.2
correspond respectively to $\eps=0$ and $\eps=1$ in our classification.

As specific subexamples, we may emphasize the generalized orbifold-string systems of cyclic permutation-type:
\begin{equation}\lb{10.3}
H_+ \subset \bigl\lbrace \Z_K \times (\thickone)_d \times H({\rm perm})_{26-d}^{'}; \,
\Z_K \times (-\thickone)_d \times H({\rm perm})_{26-d}^{'} \bigr\rbrace.
\end{equation}
In these cases, the $j$-cycles of the $K-1$ non-trivial elements of $H({\rm perm})_K=\Z_K$ have all possible lengths $\fjs \geqslant 2$ which divide K
\begin{equation}\lb{10.4}
j=0,1,\dots,\tfrac{K}{\rho \ps}-1, \quad \hj=0,1,\dots ,\fjs-1
\end{equation}
where $\rho\ps$ is the order of $\omega \ps \in \Z_K$, and correspondingly, the trivial element of $\Z_K$ has $\fjs=1$, $j=0,1,\dots,K-1$.

For all possible choices of the divisor (10.3), our discussion above then provides the following schemata
%manu10.3

\begin{equation*}
\setlength{\extrarowheight}{0.5em}
\begin{tabular}{| c | c | c | c |}
	\hline
	$\Z_K $ & $ \{ \fjs \}$ & $\eps=0$ & $\eps =1$\\[0.5em]
	\hline\hline
	$K=2^n $	& $\{$even$\}$ &  $L$ & $L$\\[0.5em]
	\cline{2-4}
	& $\fjs=1$ (trivial) & $L$ & $E$\\[0.5em]
	\hline
	$K$ odd	 & $\{$odd$\}$ &  $L$ & $E$\\[0.5em]
	\hline
	$K$ even $\neq 2^n$ & $\{$even$\}$ &  $L$ & $L$\\[0.5em]
	\cline{2-4}
	& $\{$odd$\}$ & $L$ & $E$\\[0.5em]
	\hline
\end{tabular}
\end{equation*}

%manu 10.4
\vspace{.1in}
\noindent where $L$ or $E$ denote $j$-cycles with respectively Lorentzian 
or Euclidean target space-times. The dimensionalities of these 
space-times are as follows 
\begin{equation}\lb{10.5}
\djs=
\begin{cases}
d+\djs^{(26-d)} \quad {\rm for}\; L\\[0.5em]
\djs^{(26-d)} \quad {\rm for} \;E
\end{cases}
\end{equation}
and general formulae for the contributions $\djs^{(26-d)}$ of type $(26-d)$ are given in Sec.~5.

In fact 
the results for  $\djs^{(26-d)}$ and the target space-time dimensionalities in Eq.~(10.5) apply more generally
to all $j$-cycles in any sector $\sigma$ of the large example (10.1) of orbifold-string
systems of permutation-type, with $SO(\djs-1,1)$ target space-time symmetry for Lorentzian cycles
and $SO(\djs^{(26-d)})$ target space-time symmetry for Euclidean cycles.

The Lorentzian cycles appear so far to be new physical generalizations of ordinary string 
theory, with $D_{j}(\sigma)$-dimensional ordinary Lorentzian string 
subsectors and $26-D_{j}(\sigma)$ extra characteristically-twisted scalar 
fields. The Euclidean cycles are more exotic, being comprised at most of short collections 
of particles with $-2\leq P_{j}^{2}(\sigma)\leq 0$ -- some possibly with negative-norm physical states. On the 
other hand, many of the Euclidean cycles are \emph {extinguished}, with no physical 
particles at all, 
and there is even a subset of extinguished cycles which we have called 
\emph {null} (no target space-time dimensions and no physical particles).

Although obtained explicitly as \emph {enhanced} target space-time symmetries in Secs.~7 and 8,
we emphasize that these Lorentzian and Euclidean symmetries are completely natural for the (similarly-enhanced) target space-time 
dimensionalities in Eq.~(10.5).  One therefore expects these same natural space-time 
symmetries for all the target space-times [2] of the general 
orbifold-string theories in Eq.~(1.1). One also expects the 
as-yet-unconstructed twist-fields (intertwiners) of these orbifolds to 
describe transitions among the various target space-time dimensionalities, 
signatures and symmetries of each theory.

We focus finally on orbifold-string systems in our large example whose 
target space-times are \emph{entirely Lorentzian}, for all cycles $j$ in all
sectors $\sigma$ of the orbifold.  Starting with the cyclic examples in 
Eq.~(10.3) and the Table above, one simple subset of entirely Lorentzian theories  has the divisors
\begin{equation}\lb{10.6}
H_+ = \Z_K \times (\thickone)_d \times H({\rm perm})_{26-d}^{'}
\end{equation}
where we have kept only the subgroup of $H_{+}$ with $\eps=0$.  Indeed, we have  seen in Sec.~7 that \emph{all} the 
generalized orbifold-string systems with $\eps=0$ 
\begin{equation}\lb{10.7}
H_+ =H({\rm perm})_{K} \times (\thickone)_d \times H({\rm perm})_{26-d}^{'}
\end{equation}
consist entirely of Lorentzian cycles.

%manu10.5
Again consulting the Table above, we can find many other entirely Lorentzian orbifold-string systems  in 
the cyclic subexamples (10.3). This includes the following simple cases [5,7] with $H({\rm perm})_2 = \Z_2$: 
\begin{subequations}\lb{10.8}
\begin{gather}
\lb{10.8a}
\frac{U(1)^{52}}{H_+}, \quad \left[\frac{{\rm U}(1)^{52}}{H_+}\right]_{{\rm open}}\\[0.5em]
\lb{10.8b}
H_+=\{ \tau_0 \times (\thickone)_d \times H({\rm perm})_{26-d}^{'}; \, \tau_+\times (-\thickone)_d \times H({\rm perm})_{26-d}^{'} \}.
\end{gather}
\end{subequations}
Here $(\tau_0=\thickone, \tau_+) \in \Z_2$ and the elements of $H_+$ with 
$\eps=1$ are listed after the semi-colon. 
%Here we have been careful not 
%to associate the trivial element of $$ to the non-trivial element
%to include the unit element of $$ when 

%manu 10.6
The entirely Lorentzian systems in Eq.~(10.8) are closely related to the 
orientation-orbifold string systems [19,20,22,3-6]
\begin{subequations}\lb{10.9}
\begin{gather}
\lb{10.9a}
\frac{U(1)^{26}}{H_-}=\frac{\uts_L \times \uts_R}{H_-}\\[0.5em]
\lb{10.9b}
H_-=\{ \tau_0 \times (\thickone)_d \times H({\rm perm})_{26-d}^{'}; \, \tau_-\times (-\thickone)_d \times H({\rm perm})_{26-d}^{'} \}
\end{gather}
\end{subequations}
where $(\tau_0=1, \tau_-)$ are the elements of the world-sheet orientation-reversing $\Z_2$ called $\Z_2({\rm w.s.})$.  
%manu 10.7
We remind that the orientation-orbifold string systems are the natural 
generalization of orientifolds, each system containing in fact an equal
number of twisted closed strings and twisted open strings. The twisted 
closed- and open-string sectors of these systems are given respectively before 
and after the semicolon in Eq.~(10.9b), with central charges:
\begin{subequations}\lb{10.10}
\begin{gather}
\lb{10.10a}
%{\rm closed}:\quad f(\sigma) = 1,\quad \hat{c}(\sigma)\, = \,c(\sigma) = 26\\
{\rm closed}:\quad \hat{c}_{0}(\sigma)=26, \; f_{0}(\sigma)=1, \;j=\hat{j}=0,\; c_{0}(\sigma)=26,\\
\lb{10.10b}
{\rm open}: \quad \hat{c}_{0}(\sigma)=52, \; f_{0}\ps=2, \; \bhj=0,1,\; c_{0}\ps=26.
\end{gather}
\end{subequations}
Indeed the ordinary critical bosonic open-closed string system has been identified as the simple case
\begin{equation}\lb{10.11}
H_-=\{ \tau_0 \times (\thickone)_{26};\,  \tau_- \times (-\thickone)_{26} \}
\end{equation}
even at the interacting level [6].
%manu 10.7'

More precisely, our computational results above hold as well for all the 
orientation-orbifold string systems, 
so all the sectors of all these systems are Lorentzian. The set of all
closed-string sectors of each of these systems is the ordinary space-time orbifold
\begin{equation}\lb{10.12}
\frac{U(1)^{26}}{H_1}, \quad H_1= (\thickone)_d \times H({\rm perm})_{26-d}^{'}
\end{equation}
which is described in the formulation of this paper as sets of sectors with $f_{j}(\sigma)=1$.  In such cases of course the
unreduced and reduced formulations are the same, with all sectors at central charge 26. 
%manu 10.8
The twisted open-string sectors of any orientation-orbifold system are in fact contained in the 
open-string analogues $[U(1)^{52}/H_+]_{{\rm open}}$ described here at 
$\fjs=2$, and we find from the results (5.17) and (10.5) of the text that the dimension of the Lorentzian space-time of each twisted open-string sector is 
\begin{equation}\lb{10.13}
D\ps=d+N_O\ps'+2N_E\ps'
\end{equation}
where $N_{O,E}\ps'$ are respectively the number of odd and even $L$-cycles in the 
chosen element of $H({\rm perm})_{26-d}^{'}$.

%manu 10.8'
%As discussed more generally above, each Lorentzian cycle $j$ of each sector $\sigma$ of any  orbifold-string system
%contains an ordinary untwisted $\djs$-dimensional Lorentzian string subsysystem, plus extra $SO(\djs-1,1)$ scalar contributions.  So
Because the Lorentzian cycles described in this paper have a 
$\djs$-dimensional ordinary-string subspace,
the next challenge for the new string theories is to prove the necessary 
condition for ghost-decoupling [24]
\begin{equation}\lb{10.14}
\djs\leqslant 26, \quad \forall j,\sigma
\end{equation}
in each cycle $j$ of all Lorentzian orbifold-string systems of 
permutation-type. At this point the upper bound (10.14) is only a 
strongly-motivated conjecture [3], which has so far been established only for the pure 
permutation orbifolds [2] with trivial $H'_{26}$. In a 
subsequent paper [25], we shall employ in particular the result in Eq.~(10.13) to show that the
upper bound (10.14) holds for all the cycles of the large example (10.9) 
of orientation-orbifold string systems !
%with $H({\rm perm})_2=\Z_2({\rm w.s.})$ and any $H({\rm perm})_{26-d}^{'}$ !
%manu 10.8'
%We remind that ghost-decoupling has been conjectured [3] on physical grounds 
%for all the orbifold-string theories of permutation-type, and the results of Sec.~5 will be useful in a more systematic investigation 
%of the necessary Lorentzian upper bound in Eq.~(10.14).

%manu 10.9
%As also seen in the table above, it is not difficult to construct orbifold-string systems which contain
%at least some \emph{Euclidean cycles} $(P^2_j\ps \leqslant 0)$.  Such 
%cycles (see Secs.~ 8 and 9) are rather more exotic than the Lorentzian cycles: Euclidean cycles do not contain
%ordinary string subsystems, and consist at most of "short" sets of 
%tachyonic particles with possible
 %zero-mass particles at the top of the short set. Although we intend to further study orbifold systems
%with Euclidean cycles elsewhere, we have confined ourselves here only to some simple examples (see Sec.~9) with null (extinguished) Euclidean
%cycles, which have no space-time dimensions $(\djs=0)$ and no physical particles at all.

%\clearpage
\section*{Acknowledgments}
For helpful discussion and encouragement, I thank L. Alvarez Gaum\'{e}, C. Bachas, J. de Boer, S. Frolov, O. Ganor, E. Kiritsis, A. Neveu, H. Nicolai,
N. Obers, B. Pioline, M. Porrati, E. Rabinovici,  V. Schomerus, C. Schweigert, M.  Staudacher, R. Stora, C. Thorn, E. Verlinde and J.-B. Zuber.
%\clearpage

%\bibliographystyle{annals-alpha}
%\bibliography{OstIII}

\end{document}